\documentclass[twocolumn]{aastex62}
\usepackage[utf8]{inputenc}
\usepackage{amsmath}
\usepackage{CJKutf8}

\usepackage[normalem]{ulem}
\usepackage{color}
\usepackage{graphicx}
\definecolor{red}{rgb}{1.0,0.0,0.0}

\shorttitle{HR 8799 Dynamical Constraints} 
\shortauthors{Wang et al.}

\turnoffediting

\begin{document}
\title{Dynamical Constraints on the HR 8799 Planets with GPI}

\correspondingauthor{Jason J. Wang}
\email{j-wang@berkeley.edu}
\author[0000-0003-0774-6502]{Jason J. Wang}
\affiliation{Department of Astronomy, University of California, Berkeley, CA 94720, USA}

\author{James R. Graham}
\affiliation{Department of Astronomy, University of California, Berkeley, CA 94720, USA}

\author[0000-0001-9677-1296]{Rebekah Dawson}
\affiliation{Department of Astronomy \& Astrophysics, Center for Exoplanets and Habitable Worlds, The Pennsylvania State University, University Park, PA 16802}

\author{Daniel Fabrycky}
\affiliation{Department of Astronomy and Astrophysics, University of Chicago, 5640 South Ellis Avenue, Chicago, IL 60637, USA}

\author[0000-0002-4918-0247]{Robert J. De Rosa}
\affiliation{Department of Astronomy, University of California, Berkeley, CA 94720, USA}

\author{Laurent Pueyo}
\affiliation{Space Telescope Science Institute, Baltimore, MD 21218, USA}

\author[0000-0002-9936-6285]{Quinn Konopacky}
\affiliation{Center for Astrophysics and Space Science, University of California San Diego, La Jolla, CA 92093, USA}

\author[0000-0003-1212-7538]{Bruce Macintosh}
\affiliation{Kavli Institute for Particle Astrophysics and Cosmology, Stanford University, Stanford, CA 94305, USA}

\author[0000-0002-4164-4182]{Christian Marois}
\affiliation{National Research Council of Canada Herzberg, 5071 West Saanich Rd, Victoria, BC, V9E 2E7, Canada}
\affiliation{University of Victoria, 3800 Finnerty Rd, Victoria, BC, V8P 5C2, Canada}

\author{Eugene Chiang}
\affiliation{Department of Astronomy, University of California, Berkeley, CA 94720, USA}

\author[0000-0001-5172-7902]{S. Mark Ammons}
\affiliation{Lawrence Livermore National Laboratory, Livermore, CA 94551, USA}

\author[0000-0001-6364-2834]{Pauline Arriaga}
\affiliation{Department of Physics \& Astronomy, University of California, Los Angeles, CA 90095, USA}

\author[0000-0002-5407-2806]{Vanessa P. Bailey}
\affiliation{Jet Propulsion Laboratory, California Institute of Technology, Pasadena, CA 91109, USA}

\author[0000-0002-7129-3002]{Travis Barman}
\affiliation{Lunar and Planetary Laboratory, University of Arizona, Tucson AZ 85721, USA}

\author{Joanna Bulger}
\affiliation{Subaru Telescope, NAOJ, 650 North A{'o}hoku Place, Hilo, HI 96720, USA}

\author[0000-0001-6305-7272]{Jeffrey Chilcote}
\affiliation{Department of Physics, University of Notre Dame, 225 Nieuwland Science Hall, Notre Dame, IN, 46556, USA}

\author[0000-0003-0156-3019]{Tara Cotten}
\affiliation{Department of Physics and Astronomy, University of Georgia, Athens, GA 30602, USA}

\author{Rene Doyon}
\affiliation{Institut de Recherche sur les Exoplan{\`e}tes, D{\'e}partement de Physique, Universit{\'e} de Montr{\'e}al, Montr{\'e}al QC, H3C 3J7, Canada}

\author[0000-0002-5092-6464]{Gaspard Duch\^ene}
\affiliation{Department of Astronomy, University of California, Berkeley, CA 94720, USA}
\affiliation{Univ. Grenoble Alpes/CNRS, IPAG, F-38000 Grenoble, France}

\author[0000-0002-0792-3719]{Thomas M. Esposito}
\affiliation{Department of Astronomy, University of California, Berkeley, CA 94720, USA}

\author[0000-0002-0176-8973]{Michael P. Fitzgerald}
\affiliation{Department of Physics \& Astronomy, University of California, Los Angeles, CA 90095, USA}

\author[0000-0002-7821-0695]{Katherine B. Follette}
\affiliation{Physics and Astronomy Department, Amherst College, 21 Merrill Science Drive, Amherst, MA 01002, USA}

\author{Benjamin L. Gerard}
\affiliation{University of Victoria, 3800 Finnerty Rd, Victoria, BC, V8P 5C2, Canada}
\affiliation{National Research Council of Canada Herzberg, 5071 West Saanich Rd, Victoria, BC, V9E 2E7, Canada}

\author[0000-0002-4144-5116]{Stephen J. Goodsell}
\affiliation{Gemini Observatory, 670 N. A'ohoku Place, Hilo, HI 96720, USA}

\author[0000-0002-7162-8036]{Alexandra Z. Greenbaum}
\affiliation{Department of Astronomy, University of Michigan, Ann Arbor, MI 48109, USA}

\author[0000-0003-3726-5494]{Pascale Hibon}
\affiliation{European Southern Observatory, Alonso de Cordova 3107, Vitacura, Santiago, Chile}

\author[0000-0003-1498-6088]{Li-Wei Hung}
\affiliation{Natural Sounds and Night Skies Division, National Park Service, Fort Collins, CO 80525, USA}

\author{Patrick Ingraham}
\affiliation{Large Synoptic Survey Telescope, 950N Cherry Ave., Tucson, AZ 85719, USA}

\author{Paul Kalas}
\affiliation{Department of Astronomy, University of California, Berkeley, CA 94720, USA}
\affiliation{SETI Institute, Carl Sagan Center, 189 Bernardo Ave.,  Mountain View CA 94043, USA}

\author{James E. Larkin}
\affiliation{Department of Physics \& Astronomy, University of California, Los Angeles, CA 90095, USA}

\author{J\'er\^ome Maire}
\affiliation{Center for Astrophysics and Space Science, University of California San Diego, La Jolla, CA 92093, USA}

\author[0000-0001-7016-7277]{Franck Marchis}
\affiliation{SETI Institute, Carl Sagan Center, 189 Bernardo Ave.,  Mountain View CA 94043, USA}

\author[0000-0002-5251-2943]{Mark S. Marley}
\affiliation{NASA Ames Research Center, Mountain View, CA 94035, USA}

\author[0000-0003-3050-8203]{Stanimir Metchev}
\affiliation{Department of Physics and Astronomy, Centre for Planetary Science and Exploration, The University of Western Ontario, London, ON N6A 3K7, Canada}
\affiliation{Department of Physics and Astronomy, Stony Brook University, Stony Brook, NY 11794-3800, USA}

\author[0000-0001-6205-9233]{Maxwell A. Millar-Blanchaer}
\altaffiliation{NASA Hubble Fellow}
\affiliation{Jet Propulsion Laboratory, California Institute of Technology, Pasadena, CA 91109, USA}

\author[0000-0001-6975-9056]{Eric L. Nielsen}
\affiliation{Kavli Institute for Particle Astrophysics and Cosmology, Stanford University, Stanford, CA 94305, USA}

\author[0000-0001-7130-7681]{Rebecca Oppenheimer}
\affiliation{Department of Astrophysics, American Museum of Natural History, New York, NY 10024, USA}

\author{David Palmer}
\affiliation{Lawrence Livermore National Laboratory, Livermore, CA 94551, USA}

\author{Jennifer Patience}
\affiliation{School of Earth and Space Exploration, Arizona State University, PO Box 871404, Tempe, AZ 85287, USA}

\author[0000-0002-3191-8151]{Marshall Perrin}
\affiliation{Space Telescope Science Institute, Baltimore, MD 21218, USA}

\author{Lisa Poyneer}
\affiliation{Lawrence Livermore National Laboratory, Livermore, CA 94551, USA}

\author[0000-0002-9246-5467]{Abhijith Rajan}
\affiliation{Space Telescope Science Institute, Baltimore, MD 21218, USA}

\author[0000-0003-0029-0258]{Julien Rameau}
\affiliation{Institut de Recherche sur les Exoplan{\`e}tes, D{\'e}partement de Physique, Universit{\'e} de Montr{\'e}al, Montr{\'e}al QC, H3C 3J7, Canada}

\author[0000-0002-9667-2244]{Fredrik T. Rantakyr\"o}
\affiliation{Gemini Observatory, Casilla 603, La Serena, Chile}

\author[0000-0003-2233-4821]{Jean-Baptiste Ruffio}
\affiliation{Kavli Institute for Particle Astrophysics and Cosmology, Stanford University, Stanford, CA 94305, USA}

\author[0000-0002-8711-7206]{Dmitry Savransky}
\affiliation{Sibley School of Mechanical and Aerospace Engineering, Cornell University, Ithaca, NY 14853, USA}

\author{Adam C. Schneider}
\affiliation{School of Earth and Space Exploration, Arizona State University, PO Box 871404, Tempe, AZ 85287, USA}

\author[0000-0003-1251-4124]{Anand Sivaramakrishnan}
\affiliation{Space Telescope Science Institute, Baltimore, MD 21218, USA}

\author[0000-0002-5815-7372]{Inseok Song}
\affiliation{Department of Physics and Astronomy, University of Georgia, Athens, GA 30602, USA}

\author[0000-0003-2753-2819]{Remi Soummer}
\affiliation{Space Telescope Science Institute, Baltimore, MD 21218, USA}

\author{Sandrine Thomas}
\affiliation{Large Synoptic Survey Telescope, 950N Cherry Ave., Tucson, AZ 85719, USA}

\author{J. Kent Wallace}
\affiliation{Jet Propulsion Laboratory, California Institute of Technology, Pasadena, CA 91109, USA}

\author[0000-0002-4479-8291]{Kimberly Ward-Duong}
\affiliation{School of Earth and Space Exploration, Arizona State University, PO Box 871404, Tempe, AZ 85287, USA}

\author{Sloane Wiktorowicz}
\affiliation{Department of Astronomy, UC Santa Cruz, 1156 High St., Santa Cruz, CA 95064, USA }

\author[0000-0002-9977-8255]{Schuyler Wolff}
\affiliation{Leiden Observatory, Leiden University, P.O. Box 9513, 2300 RA Leiden, The Netherlands}


\begin{abstract}
The HR 8799 system uniquely harbors four young super-Jupiters whose orbits can provide insights into the system's dynamical history and constrain the masses of the planets themselves. Using the Gemini Planet Imager (GPI), we obtained down to one milliarcsecond precision on the astrometry of these planets. We assessed four-planet orbit models with different levels of
constraints and found that assuming the planets are near 1:2:4:8 period commensurabilities, or are coplanar, does not worsen the fit. We added the prior that the planets must have been stable for the age of the system (40~Myr) by running orbit configurations from our posteriors through $N$-body simulations and varying the masses of the planets. We found that only assuming the planets are both coplanar and near 1:2:4:8 period commensurabilities produces dynamically stable orbits in large quantities. Our posterior of stable coplanar orbits tightly constrains the planets' orbits, and we discuss implications for the outermost planet b shaping the debris disk. 
A four-planet resonance lock is not necessary for stability up to now. However, planet pairs d and e, and c and d, are each likely locked in two-body resonances for stability if their component masses are above $6~M_{\rm{Jup}}$ and $7~M_{\rm{Jup}}$, respectively.
Combining the dynamical and luminosity constraints on the masses using hot-start evolutionary models and a system age of $42 \pm 5$~Myr, we found the mass of planet b to be $5.8 \pm 0.5~M_{\rm{Jup}}$, and the masses of planets c, d, and e to be $7.2_{-0.7}^{+0.6}~M_{\rm{Jup}}$ each.
\end{abstract}

\section{Introduction}
High-contrast imaging spatially separates the faint light of planets from the bright glare of their host star. By monitoring exoplanetary systems with high-contrast imaging, we are able to obtain footage of these exoplanets in motion and trace out their orbits. Orbit analysis has been a powerful tool in characterizing the dynamics of directly-imaged systems. Through orbital monitoring of $\beta$~Pic~b, we now know that the planet is responsible for inducing the observed warp in the circumstellar debris disk \citep{Dawson2011,Lagrange2012}, although it may not be alone in clearing out the cavity of the disk \citep{MillarBlanchaer2015}. Precise orbital determination also has timed the Hill sphere transit of the planet to between April of 2017 to January of 2018 \citep{Wang2016}, which offered a unique opportunity to probe the circumplanetary environment of a young exoplanet \citep{Stuik2017,Mekarnia2017,deMooij2017}. 
For HD~95086~b, by combining orbit fits with constraints on the debris disk geometry, \citet{Rameau2016} showed that the planet alone cannot be clearing out the gap in the system, and that additional planets reside closer in to the star. The orbit of Fomalhaut b was shown to cross the debris disk in the system, revealing that the planet cannot be a massive Jupiter-like planet, but rather a dwarf planet shrouded by dust \citep{Kalas2013}. Finally, future orbital monitoring of 51~Eri~b could shed light on the interactions between the planet and the wide-separation binary GJ~3305 \citep{DeRosa2015}. 

Long-term orbital monitoring can also lead to dynamical mass measurements of the planets themselves, which will assess evolutionary models of young giant planets that all current mass estimates of directly-imaged exoplanets are based on \citep{Baraffe2003,Marley2007}. In the coming years, Gaia will measure the astrometric reflex motion of stars hosting planets \citep{Perryman2014}. Gaia astrometry combined with long-term orbital monitoring from direct imaging will provide the tightest model-independent constraints on the masses of the planets \citep{Sozzetti2016}. Alternatively, multi-planet systems where planets mutually perturb their orbits provide another way to constrain the masses of the planets in the system. In resonant systems where the dynamical timescales are close to the orbital timescales, such mutual perturbations have been measured in short period planets as variations in the host star's radial velocity signature \citep[e.g.,][]{Marcy2001,Rivera2010} and as transit timing variations \citep[e.g.,][]{Agol2005,Holman2005}, leading to direct measurements of the masses. Due to the long orbital periods of known directly-imaged systems, such a direct measurement of the mutual perturbations on the orbits has been impossible with the current observational baselines, none of which span a full orbital period. Still, upper limits on the masses of the planets based on dynamical stability can be obtained. 
Stability mass constraints have been used to characterize exoplanets discovered in compact systems, such as TRAPPIST-1 (e.g., \citealt{Quarles2017,Tamayo2017}), Kepler-36 \citep{Deck2012}, and the HR 8799 system discussed in this paper.

HR 8799 is unique among directly-imaged systems as it is the only one known to harbor four planets \citep{Marois2008,Marois2010}. 
The planets orbit $\sim$15-70~au from the star between two rings of rocky bodies, similar to the configuration of the giant planets in our own Solar System \citep{Su2009}. The outer belt has been resolved with far-infrared and millimeter observations, although the exact orientation and inner edge of the disk are not entirely agreed upon \citep{Hughes2011,Matthews2014,Booth2016,Wilner2018}. Assuming ``hot-start" evolutionary models and an age of 30~Myr, \citet{Marois2008,Marois2010} translated the planet luminosities into masses: planet b is $\sim$5~$M_{\rm{Jup}}$ while the inner three planets are $\sim$7~$M_{\rm{Jup}}$ \citep{Marois2008,Marois2010}. However, as the evolutionary models are uncertain at these early ages, so are the exact masses of the planets. Fortunately, dynamics can provide an additional constraint on the masses of the planets, even if their long orbital periods mean we cannot detect planet-planet interactions and fully constrain the masses this way. 

Since the discovery of the HR 8799 planets, their orbits have been closely monitored. Keplerian orbits have been fit to the astrometry obtained from many instruments using least-squares techniques that look for families of orbits or Bayesian parameter estimation with Markov-chain Monte Carlo (MCMC) methods that explore the full posterior of orbital parameters \citep{Soummer2011b,Currie2012,Esposito2013,Maire2015,Pueyo2015,Zurlo2016,Konopacky2016,Wertz2017}. Fitting the planets independently, some studies have reported planet d to be misaligned in its orbit relative to the other planets \citep{Currie2012,Esposito2013,Pueyo2015} or one of the inner planets having eccentricities above 0.2 \citep{Maire2015,Wertz2017}. However, several of the authors have noted that unaccounted astrometric calibration offsets between instruments may be inducing inclination and eccentricity biases \citep{Pueyo2015,Maire2015,Konopacky2016}. Recently, \citet{Konopacky2016} presented self-consistent astrometry using only measurements from Keck and found that coplanar and low-eccentricity solutions were consistent with the data. Despite the uniform analysis, the 7~years of Keck data still only cover a short arc of these orbits that have periods between $\sim$40-400~years, leaving many possible orbital configurations.

The measured astrometry is not the only constraint on the orbit of these planets. HR 8799 is part of the Columba moving group \citep{Zuckerman2011}, a group of stars that formed together  $42^{+6}_{-4}$~Myr ago \citep{Bell2015}. Thus the four planets need to be stable dynamically for almost the same amount of time. 
Studies using $N$-body simulations have explored the dynamical constraints on the orbital parameters and masses. These studies have found stable orbits using the nominal luminosity-derived masses from \citet{Marois2010} without invoking orbital resonances \citep{Sudol2012,Gotberg2016} or to even higher masses assuming long-term resonance lock of the planets \citep{Fabrycky2010,Marois2010,Gozd2014,Gozd2018}. However, many of these studies initialize or fit the simulated orbits to one astrometric measurement, leaving a gap between orbit fits from the data and dynamical constraints from simulations \citep{Fabrycky2010,Marois2010,Sudol2012,Gotberg2016}. To connect simulations to the data more rigorously, \citet{Gozd2014} developed a novel technique to lock the planets into resonance and then search for times and orientations that matched all of the available data. Their orbit and mass constraints though only apply to the family of orbits that slowly migrated into a four-planet resonance lock. 

A few attempts have been made to include stability in the orbit fitting of this system.
Analytical prescriptions have been used to remove the orbits that are most obviously not dynamically stable \citep{Pueyo2015,Konopacky2016}. \citet{Esposito2013} ran $N$-body simulations on their orbital fits from a least-squares algorithm and only found stable orbits up to $5~M_{\rm{Jup}}$. In general, finding stable orbits in the orbit fits has been impractical with short orbital arcs. Having only the 2-D sky projection of an arc of an orbit, even with milliarcsecond-level precision, cannot break many degeneracies in the orbital parameters resulting in too wide a variety of orbital solutions which are nearly all unstable.

In this paper, we present an analysis that better bridges the gap between orbit fits and dynamical constraints by incorporating $N$-body simulations as a rejection sampling step of our Bayesian orbit fit to enforce stability.
In Section \ref{sec:obs}, we show we have obtained one milliarcsecond astrometry of all four planets using the Gemini Planet Imager \citep[GPI;][]{Macintosh2014} and the open-source \texttt{pyKLIP} data reduction package \citep{Wang2015}. In Section \ref{sec:orbitfit}, we combine the precise GPI measurements with the uniformly-reduced Keck astrometry measured by \citet{Konopacky2016} and fit multiple orbital models with different assumptions about coplanarity and resonance using MCMC techniques that sample the full posterior of possible orbital configurations. 
In Section \ref{sec:dynamics}, we take the posteriors of orbits from our Bayesian analysis and simulate them for 40~Myr using 
the \texttt{REBOUND} $N$-body integrator \citep{Rein2012} to find the posterior of stable orbits after applying a dynamical stability prior. We discuss the consequences of our results, such as planets shaping the cold debris disk, the necessity of orbital resonances for stability, dynamical limits on the masses of the planets, and the future stability of the system.

\section{Observations and Data Reduction}\label{sec:obs}

\begin{deluxetable}{cccccc}
\tabletypesize{\footnotesize}
\tablewidth{0pt} 
\tablecaption{GPI Observations of HR 8799 \label{table:obs} }
\tablehead{ 
 UT Date & \shortstack[c]{Filter} & \shortstack[c]{ \\ \\Exposure\\Time (s)} & \shortstack[c]{Field\\Rotation (\degr)} & \shortstack[c]{Planets\\Imaged}
}
\startdata 
2013 Nov 17 & \textit{K1} & 2130 & 17 & cde \\
2014 Sep 12 & \textit{H} & 3107 & 19 & bcd \\
2016 Sep 19 & \textit{H} & 3579 & 21 & cde \\
\enddata
\end{deluxetable}

To obtain astrometry of the planets, we used three epochs of observations of HR 8799 taken with the integral field spectroscopy (IFS) mode of GPI. Two epochs were from instrument commissioning (Gemini program GS-ENG-GPI-COM) and one epoch from the GPI Exoplanet Survey (Gemini program GS-2015B-Q-500; PI: Macintosh). Details of the three observations are listed in Table \ref{table:obs}. While HR 8799 b is normally located outside of the field of view of GPI, we steered the field of view on the detector during the 2014 September 12 observations to see planet b, although the conditions in this dataset were too poor to see planet e. 

Raw IFS data from each epoch were processed to create 3-D spectral datacubes using the automated data reduction system for the GPI Exoplanet Survey \citep{Wang2018}. Briefly, the data were dark subtracted, individual micro-spectra on the detector were extracted to form spectral datacubes, bad pixels were corrected, distortion in the image was corrected, and satellite spots, fiducial diffraction spots centered about the location of the star, were located. The star center in each wavelength channel is estimated using the satellite spots to correct any remaining differential atmospheric refraction not removed by the atmospheric dispersion corrector. See Appendix A of \citet{Wang2018} for details.

\begin{deluxetable*}{cccccc}
\tablecaption{Astrometric Measurements of the HR 8799 planets \label{table:astrom} }
\tablewidth{0pt} 
\tablehead{ 
UT date  &  Planet  &  \shortstack[c]{KL Modes}  &  \shortstack[c]{\\ \\ Exclusion \\ Criterion (pixels)} & \shortstack[c]{Radial \\ Separation (mas) } & \shortstack[c]{Position Angle \\  (\degr)}
}
\startdata
2013 Nov 17 & c & 10 & 3  & 949.5 $\pm$ 0.9 & 325.18 $\pm$ 0.14 \\
 & d & 10 & 3  & 654.6 $\pm$ 0.9 & 214.15 $\pm$ 0.15 \\
 & e & 20 & 1.5  & 382.6 $\pm$ 2.1 & 265.13 $\pm$ 0.24 \\
2014 Sep 12 & b & 10 & 1.5  & 1721.2 $\pm$ 1.4 & 65.46 $\pm$ 0.14 \\
 & c & 10 & 1.5 & 949.0 $\pm$ 1.1 & 326.53 $\pm$ 0.14 \\
 & d & 10 & 1.5 & 662.5 $\pm$ 1.3 & 216.57 $\pm$ 0.17 \\
2016 Sep 19 & c & 10 & 2 & 944.2 $\pm$ 1.0 & 330.01 $\pm$ 0.14 \\
 & d & 10 & 2 & 674.5 $\pm$ 1.0 & 221.81 $\pm$ 0.15 \\
 & e & 10 & 1 & 384.8 $\pm$ 1.7 & 281.68 $\pm$ 0.25
\enddata
\end{deluxetable*}

We used the Karhunen-Lo\`eve Image Projection algorithm \citep[KLIP;][]{Soummer2012,Pueyo2015} to subtract off the stellar glare and the Bayesian KLIP-FM Astrometry (BKA) technique \citep{Wang2016} to measure the astrometry of each planet. BKA forward models the distortions to the planet point spread function (PSF) induced by KLIP in subtracting the stellar PSF and fits for the planet position while also accounting for the correlated noise in the image as a Gaussian process. In \citet{Wang2016}, we used this technique to obtain one milliarcsecond astrometry on $\beta$~Pic~b. We used the KLIP and BKA implementations available in the \texttt{pyKLIP} package \citep{Wang2015} from commit \texttt{4f56e34}. For all the reductions, we first ran a high-pass filter to suppress the low spatial frequency background, constructed the instrumental PSF from the satellite spots, selected an annulus containing each planet to run KLIP on, and averaged the data in time and wavelength. To optimize the detection of each planet, we varied the number of Karhunen-Lo\`eve (KL) modes to model the stellar PSF, and the minimum number of pixels the planet needed to move in the reference images due to angular differential imaging \citep{Marois2006} and spectral differential imaging \citep{Marois2000}. We listed these parameters in Table \ref{table:astrom}. To measure the planets' astrometry, we used the \texttt{emcee} package \citep{ForemanMackey2013} to sample the posterior distribution for the location of the planet while also fitting the noise as a Gaussian process with spatial correlation described by the same Mat\'{e}rn covariance function as used in \citet{Wang2016}.  For each planet, our Markov-chain Monte Carlo sampler used 100 walkers, and each walker was run for 800 steps, with a ``burn-in" of 300 steps beforehand that corresponded to at least three autocorrelation times for any chain. We then added additional terms in our astrometric error budget in quadrature: a 0.05~pixel uncertainty in locating the central star \citep{Wang2014}; a plate scale of $14.166 \pm 0.007$~mas~lenslet$^{-1}$; and a residual North offset of $0\fdg10 \pm 0\fdg13$ \citep{DeRosa2015}. 

Our final astrometric results are listed in Table \ref{table:astrom}. We achieved down to 1~mas precision on the astrometry of planets b, c, and d. For these three planets that are further from the star, the dominant sources of uncertainty are from the location of the star and the astrometric calibration of GPI. We achieved 1-2~mas precision on planet e, which is limited by the signal to noise ratio of the planet. This is 1.5 to 2 times more precise than the SPHERE astrometry from \citet{Wertz2017} and at least 3 times more precise than the Keck astrometry from \citet{Konopacky2016}. 

\section{Orbit Fitting}\label{sec:orbitfit}

To investigate the possible orbital solutions for the HR~8799 planets, we combined our GPI measurements with those from Keck that were reported in \citet{Konopacky2016}. We chose to consider only these two datasets to minimize unknown systematic errors in the astrometric calibration across instruments. Specifically, GPI is astrometrically calibrated against the NIRC2 instrument at Keck, the same instrument used for the Keck HR~8799 observations, so systematic offsets between the two datasets are minimized \citep{Konopacky2014, DeRosa2015}. While \textit{Hubble Space Telescope} data from 1998 provides an additional 6~years of baseline, the 20-30~mas $1\sigma$ uncertainties are not particularly constraining so we did not use them \citep{Soummer2011b}.

\begin{deluxetable*}{cc|c|c|c|c}
\tablecaption{Orbital Parameters of HR 8799 bcde from Different Models  \label{table:mcmcfits} }
\tablewidth{0pt} 
\tablehead{ 
Body  &  \shortstack[c]{\\ \\ Orbital\\Element}  &  {\shortstack[c]{Unconstrained}} & {Coplanar} & {Near 1:2:4:8} & {\shortstack[c]{Near 1:2:4:8 \\ Coplanar Low-$e$ }}  }
\startdata
b & $a_b$ (au)        & $69.5^{+9.3}_{-7.0}$   & $66.4^{+4.1}_{-3.6}$       & $69.4^{+3.1}_{-4.0}$    &  $69.5^{+2.6}_{-2.8}$  \\
 & $\tau_b$           & $0.54^{+0.14}_{-0.16}$ & $0.46^{+0.05}_{-0.06}$     & $0.40^{+0.11}_{-0.15}$ & $0.38^{+0.11}_{-0.12}$ \\
 & $\omega_b$ (\degr) & $92^{+30}_{-34}$       & $92 \pm 15$               & $95^{+49}_{-41}$       & $102^{+39}_{-46}$  \\
 & $\Omega_b$ (\degr) & $127^{+32}_{-93}$      & $126^{+12}_{-14}$         & $82^{+36}_{-16}$       & $78^{+13}_{-10}$  \\
 & $i_b$ (\degr)      & $29^{+7}_{-8}$         & $23 \pm 5$                & $23^{+4}_{-6}$         & $24 \pm 3$  \\
 & $e_b$              & $0.15 \pm 0.05$        & $0.15 \pm 0.06$           & $0.07^{+0.06}_{-0.05}$ & $0.05^{+0.04}_{-0.03}$  \\
c & $a_c$ (au)        & $37.6^{+2.2}_{-1.7}$   & $40.5^{+2.7}_{-1.7}$      & $41.2^{+2.3}_{-1.6}$   &  $43.3^{+1.9}_{-1.7}$  \\
 & $\tau_c$           & $0.50^{+0.10}_{-0.18}$ & $0.14^{+0.18}_{-0.11}$     & $0.13^{+0.15}_{-0.09}$ & $0.09^{+0.02}_{-0.07}$  \\
 & $\omega_c$ (\degr) & $65^{+59}_{-29}$       & $52^{+83}_{-35}$          & $48^{+62}_{-29}$       & $63^{+60}_{-29}$  \\
 & $\Omega_c$ (\degr) & $110^{+38}_{-47}$      & $126^{+12}_{-14}$         & $112^{+17}_{-26}$      & $78^{+13}_{-10}$  \\
 & $i_c$ (\degr)      & $20^{+4}_{-5}$         & $23 \pm 5$                & $21^{+3}_{-4}$         & $24 \pm 3$  \\
 & $e_c$              & $0.09 \pm 0.04$        & $0.05^{+0.05}_{-0.03}$    & $0.04^{+0.05}_{-0.03}$ & $0.03^{+0.04}_{-0.02}$  \\
d & $a_d$ (au)        & $27.7^{+2.2}_{-1.7}$   & $25.3^{+1.3}_{-1.1}$      & $25.6^{+1.2}_{-1.3}$   &  $25.6^{+1.0}_{-0.9}$  \\
 & $\tau_d$           & $0.79^{+0.07}_{-0.18}$ & $0.872^{+0.019}_{-0.016}$ & $0.85 \pm 0.03$        & $0.839 \pm 0.20$  \\
 & $\omega_d$ (\degr) & $144^{+13}_{-23}$      & $133^{+15}_{-11}$         & $148^{+22}_{-137}$     & $165^{+11}_{-157}$  \\
 & $\Omega_d$ (\degr) & $92^{+27}_{-15}$       & $126^{+12}_{-14}$         & $86^{+26}_{-16}$       & $78^{+13}_{-10}$  \\
 & $i_d$ (\degr)      & $33 \pm 4$             & $23 \pm 5$                & $23^{+5}_{-6}$         & $24 \pm 3$  \\
 & $e_d$              & $0.15 \pm 0.11$        & $0.28 \pm 0.04$           & $0.20 \pm 0.05$        & $0.18^{+0.02}_{-0.03}$  \\
e & $a_e$ (au)        & $15.3^{+1.4}_{-1.1}$   & $14.0^{+0.7}_{-0.6}$      & $15.7^{+0.06}_{-0.07}$ &  $15.4 \pm 0.06$ \\
 & $\tau_e$           & $0.71^{+0.17}_{-0.33}$ & $0.91^{+0.05}_{-0.06}$    & $0.10^{+0.15}_{-0.06}$ & $0.07^{+0.05}_{-0.04}$  \\
 & $\omega_e$ (\degr) & $100^{+27}_{-49}$       & $128^{+25}_{-18}$         & $86^{+44}_{-36}$       & $76^{+25}_{-21}$  \\
 & $\Omega_e$ (\degr) & $117 \pm 17$           & $126^{+12}_{-14}$         & $90^{+9}_{-19}$        & $78^{+13}_{-10}$  \\
 & $i_e$ (\degr)      & $31 \pm 5$             & $23 \pm 5$                & $28^{+3}_{-4}$         & $24 \pm 3$  \\
 & $e_e$              & $0.13^{+0.06}_{-0.05}$  & $0.16 \pm 0.05$           & $0.05^{+0.07}_{-0.04}$ & $0.08^{+0.03}_{-0.04}$  \\
A & Parallax (mas)    & $24.70 \pm 0.16$       & $24.60 \pm 0.56$          & $24.38 \pm 0.62$       &  $24.38^{+0.54}_{-0.53}$  \\
 & $M_{\star}$ ($M_\odot$)& $1.48^{+0.05}_{-0.04}$& $1.46^{+0.12}_{-0.11}$   & $1.42^{+0.12}_{-0.11}$ & $1.40 \pm 0.10$  \\
 \hline
 & $\chi^2_{\nu}$ & $1.01^{+0.10}_{-0.07}$ & $0.88 \pm 0.07$ & $0.95^{+0.07}_{-0.06}$ & $0.94^{+0.12}_{-0.05}$  \\
 & $\Delta BIC$ & $0^{+7}_{-6}$ & $-34 \pm 0.06$ & $-4^{+5}_{-6}$ & $-29^{+6}_{-4}$  \\
\hline
\multicolumn{2}{c|}{\shortstack[c]{Stable Orbits \\(first $10^{6}$ draws)}} & 0 & 0 & 1 & 441  \\
\enddata
\tablecomments{The quoted values for $\omega$ and $\Omega$ are wrapped to be between 0\degr~and 180\degr~so posterior percentiles describe one of the two symmetric peaks. For each parameter, the median value is reported with the superscript and subscript corresponding to the 84th and 16th percentiles of the distribution respectively. For a normal distribution, these values correspond to the mean and $1\sigma$ range. }
\end{deluxetable*}

In this section, we fit the four planet orbits to four orbital configurations with increasing constraints: first, four Keplerian orbits that share the same parallax and stellar mass (Section \ref{sec:noconstrain}); second, forcing coplanarity of the four planets (Section \ref{sec:coplanar}); third, forcing the four planets to be near 1:2:4:8 period commensurabilities but with no coplanarity constraints (Section \ref{sec:1248}); lastly, forcing both coplanarity and the periods to be near a 1:2:4:8 ratio (Section \ref{sec:1248coplanar}). The constraints are intended to tighten the parameter space around stable orbits, but we are not directly considering stable orbits in these orbit fits. Dynamical stability constraints will be added in Section \ref{sec:dynamics}.

\begin{figure*}
\epsscale{1.15}
\plotone{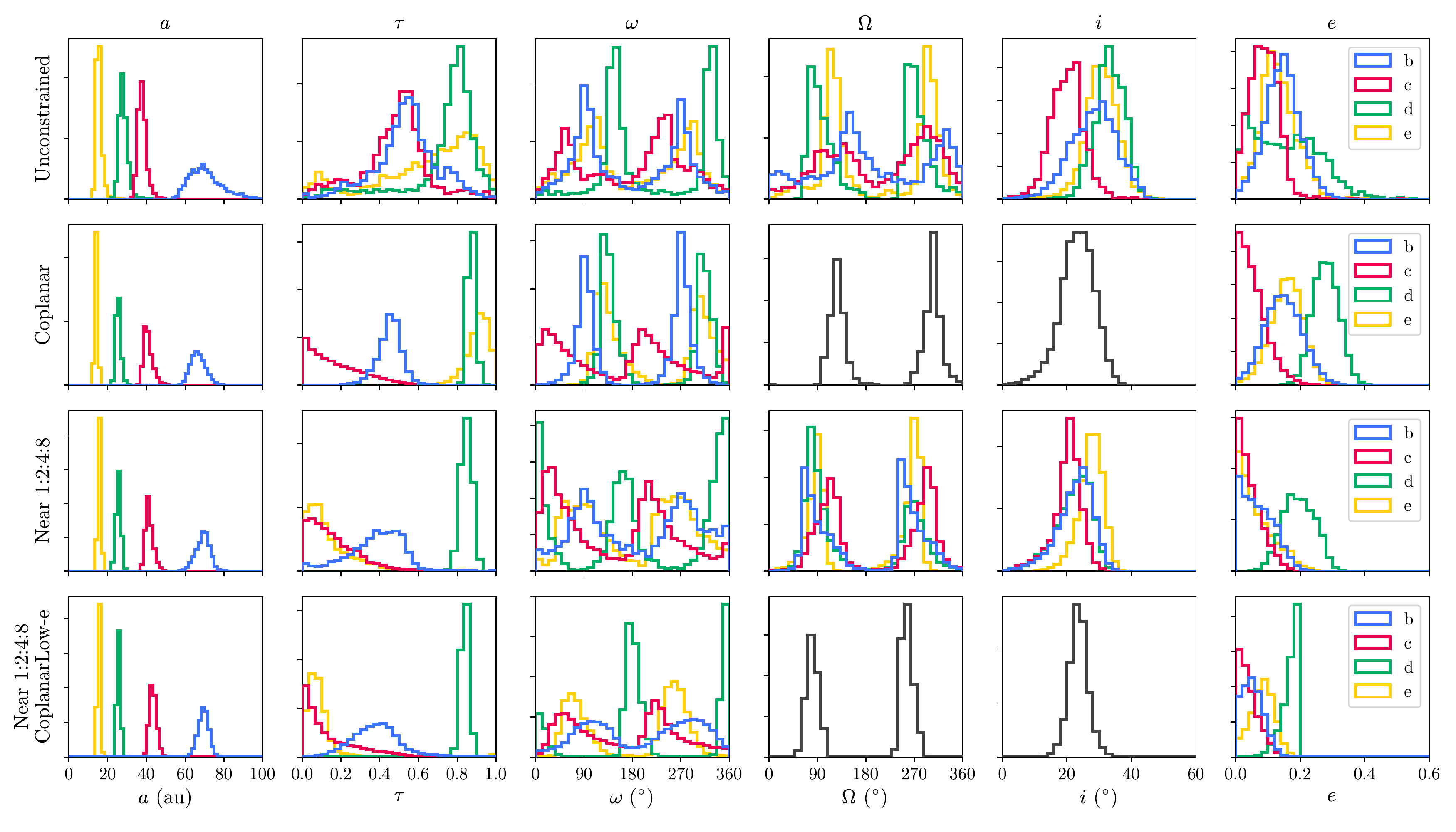}
\caption{The posteriors of each planet's orbital parameters for each of the four different models considered in Section \ref{sec:orbitfit}. Each row contains the four planet's posteriors (color coded by planet) for one model. For the coplanar models, the planets have the same $\Omega$ and $i$, so only one is plotted.
\label{fig:posteriors}}
\end{figure*}

\subsection{Unconstrained Orbits}\label{sec:noconstrain}

First, we fit four independent Keplerian orbits to the data. We employed the same Bayesian framework as \citet{Wang2016} that used Markov-chain Monte Carlo (MCMC) to sample the posterior distribution of orbital elements. For each planet, we fit for the conventional Keplerian orbital elements:
semi-major axis ($a$), epoch of periastron after MJD 50,000 in units of fractional orbital period ($\tau$), argument of periastron ($\omega$), longitude of the ascending node ($\Omega$), inclination ($i$), and eccentricity ($e$). Our conventions follow those defined in \citet{Alzner2012} for binary stars. 
In this approach each planet's orbital properties are independent, except we require that the four planets' orbits use the same parallax and total system mass, which we take to be the stellar mass.
To account for the uncertainties in the parallax and stellar mass, we assumed a Gaussian prior for the system parallax of $24.76 \pm 0.64$~mas \citep{Gaia2016} and a Gaussian prior for the stellar mass of $1.52 \pm 0.15~M_\odot$, which is mass reported by \citet{Baines2012} but with an additional 10\% uncertainty to account for systematic model errors as was done in \citet{Konopacky2016}. This case covers the full range of orbital parameters that are consistent with the data; the three following orbit fits will explore subsets of this parameter space. Due to the high dimensionality of the orbital parameters (26 in total), it will be incredibly difficult to find the dynamically stable orbits if they reside in a very small subspace. Regardless, this orbital fit is an important fiducial case to be used as a baseline model with minimal assumptions. We will refer to this orbital fit as the ``Unconstrained" fit. 

We generally used uniform priors on our orbital parameters. For each planet, the prior on $a$ was uniform in $log(a)$ between 1 and 100 au; the prior on $\tau$ was uniform between 0 and 1; the priors on $\omega$ and $\Omega$ were uniform from 0 to $2\pi$; the prior on $i$ is the geometric $\sin(i)$ prior between 0 and $\pi$; and the prior on $e$ was uniform between 0.000001 to 0.999. We note that our choice of orbital parameters will result in dual peaks in the $\omega$ and $\Omega$ posteriors that reflect our ignorance of the planets' radial velocities.

We used the parallel-tempered affine-invariant sampler \citep{Goodman2010} implemented in \texttt{emcee} \citep{ForemanMackey2013} using 15 temperatures and 1500 walkers per temperature. To improve the speed of convergence of the orbit fit, we initialized the walkers by drawing from allowed orbital parameters of individual fits to each planet using the same process. We ran each walker for 125,000 steps, after an initial burn in of 95,000 steps. Convergence was assessed using the autocorrelation time and confirming by-eye that $\omega$ and $\Omega$ had symmetric peaks. On a 32~core machine with AMD Opteron 6378 processors clocked at 2.3~GHz, this took seven days to complete, although we note that we did not make an attempt to optimize the code. We then thinned the chains by a factor of 75 to mitigate any correlation in the Markov chains. Taking only the lowest temperature walkers, we then were left with 2,499,000 samples of the posterior distribution.  The posterior distributions are plotted in Figure \ref{fig:posteriors} and reported in Table \ref{table:mcmcfits}. 

\begin{figure}[t]
\plotone{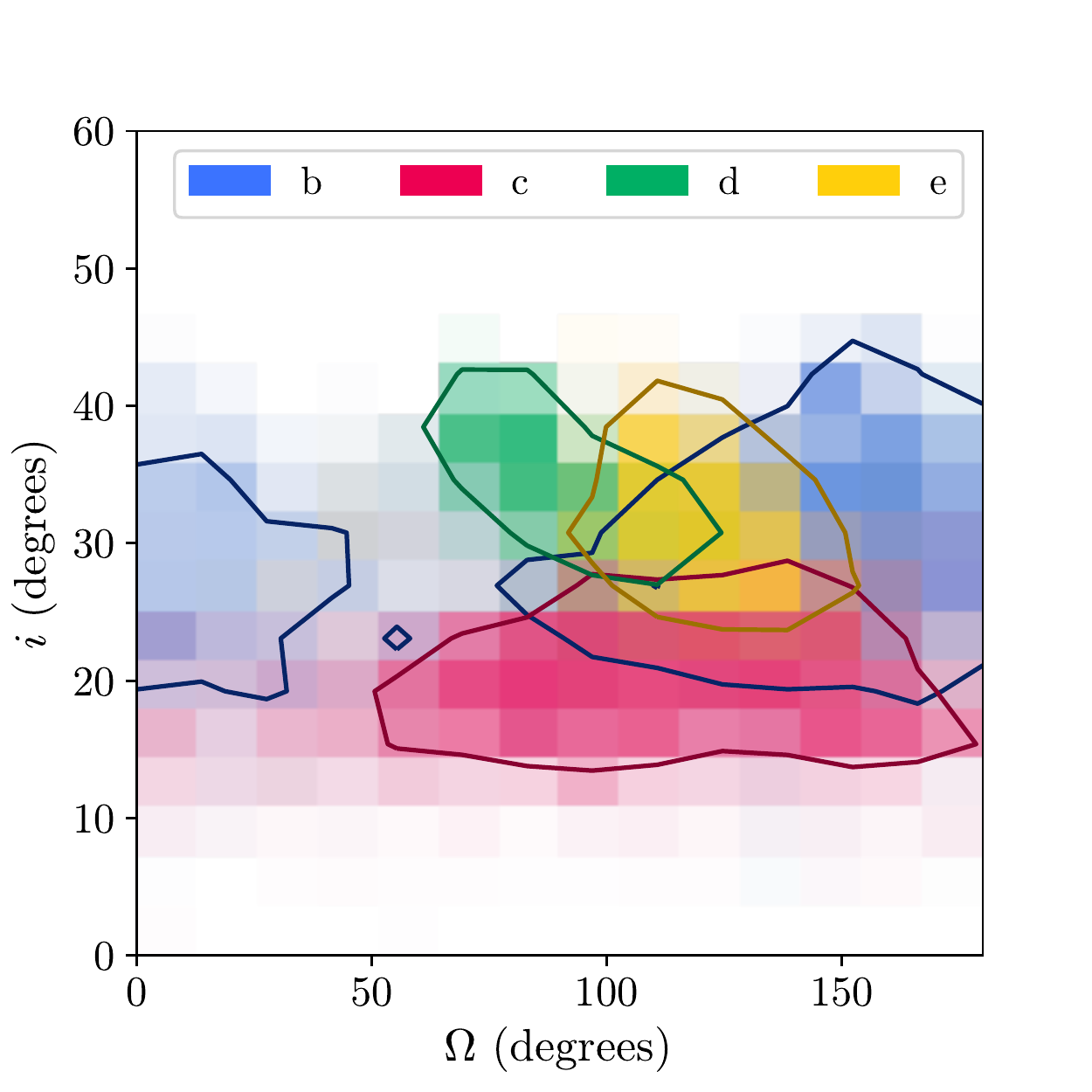}
\caption{The posteriors of each planet's angles $\Omega$ and $i$ for the Unconstrained fit. Blue, magenta, green, and yellow correspond to planets b, c, d, and e respectively. The $1\sigma$ contour is plotted on top of each planet's histogram. Overlapping regions indicate where coplanar orbits reside. Note that $\Omega$ is wrapped to only consider angles between $0^{\circ}$ and $180^{\circ}$ as the posterior is identical between $180^{\circ}$ and $360^{\circ}$ by construction.
\label{fig:planarity}}
\end{figure}

Following similar analyses from previous orbit fitting studies \citep[e.g.,][]{Konopacky2016,Wertz2017}, we investigate the mutual inclination of the planets' orbits by plotting in Figure \ref{fig:planarity} $\Omega$ and $i$, the two orbital elements that describe the orientation of the orbital plane. We will assume the planets orbit in the same direction. A planet with $\Omega$ differing by $180^{\circ}$ would be in a retrograde orbit relative to the other planets, which we do not consider here. We see that the $1\sigma$ contours for the four planets do overlap near $i\sim30^{\circ}$ and $\Omega\sim100^{\circ}$, indicating coplanar orbital solutions exist. This result agrees with the assessment of coplanarity by \citet{Konopacky2016} using similar arguments, although they preferred a different $\Omega$. 
However, we note that only 0.005\% of our sampled orbits have all four planets being mutually inclined by $< 10^{\circ}$. This result likely indicates that without any constraints on the orientation of the orbital planes, it is extremely inefficient to sample coplanar orbits in large quantities. This is not surprising since the near-coplanar solutions are just a small subset of an eight-dimensional space in which we have chosen uniform, uncorrelated priors on each parameter.   To more rigorously test coplanar orbits, we will fit directly for them (Section \ref{sec:coplanar}) and assess the fits (Section \ref{sec:goodfit}).

\subsection{Coplanar Orbits \label{sec:coplanar}}
As planets form from the circumstellar disk, it would not be surprising to find the planets 
residing in coplanar orbits. 
The posteriors from the fit without constraints are consistent with coplanarity, but does not strongly favor it. Here we will explicitly fit for coplanar orbits, and in Section \ref{sec:goodfit}, we will assess if this approach fits the data as well as the unconstrained one. 
We modify our fit so that all four planets share the same values of $\Omega$ and $i$, reducing the fit to 20 orbital parameters. We will refer to this orbital fit as the ``Coplanar" fit.  

We used a parallel-tempered sampler with 15 temperatures and 1500 walkers per temperature. We ran each walker for 87,500 steps, after an initial burn in of 132,500 steps. Convergence was assessed in the same way as in Section \ref{sec:noconstrain}. We again thinned the chains by a factor of 75, and formed our posterior distribution from the lowest temperature chains. Our posterior distribution has 1,749,000 samples. The posteriors are plotted in Figure \ref{fig:posteriors} and reported in Table \ref{table:mcmcfits}. 

From the posteriors in Figure \ref{fig:posteriors}, we see that the angles $\Omega$ and $i$ that define the orientation of the orbital plane are consistent with the orbital planes of the four planets of the Unconstrained fit. The Coplanar orbits favor inclinations between 20\degr~and 30\degr, which is $\sim$10\degr~more face-on than the solutions from \citet{Konopacky2016} with just the Keck data alone. We still find $\Omega > 90^{\circ}$, which is not preferred for coplanar orbits in \citet{Konopacky2016}. The solutions where $\Omega \approx 90^{\circ}$ which are in agreement with \citet{Konopacky2016} however favor lower inclinations near $i \approx 20^{\circ}$. While there are some differences on the preferred values, we note that many of these values are not ruled out by \citet{Konopacky2016} in their analysis.

We also find that forcing the system to be coplanar causes the eccentricity of planet d to be much higher, with $< 2\%$ of the allowed orbits having $e < 0.2$. This was due to nearly all of planet d's low eccentricity orbits from the Unconstrained fit lying outside of the range of allowed orbital planes from the Coplanar fit. As $\Omega$ and $i$ are constrained by the other three planets, raising $e_d$ provided a way to obtain the best fits to the data. We do note that the systems with $\Omega$ near $90^{\circ}$ did have the lowest eccentricities for planet d.

\subsection{Near 1:2:4:8 Period Ratio Orbits }\label{sec:1248}
We then investigated resonant orbits, focusing in particular on the 1:2:4:8 resonance, where consecutive pairs of planets are in 2:1 period resonance. We will first choose to be agnostic about the four planets' mutual inclinations. Because these planets are not massless, even if they are in resonance, they do not necessarily reside at the exact period commensurabilities. Additionally, precession of the planets' longitude of periastrons can further offset the observed period ratios from exact integer values. We note that previous orbit fitting work has assumed exact period commensurabilities when assessing if the fits were consistent with certain resonances. 

At high planet masses like the HR 8799 planets, stable period ratios for the 2:1 two-body resonance tend to be larger than 2 due to resonance overlaps at smaller period ratios causing instability \citep{Morrison2016}.
Thus, instead of fixing the period ratio of the planet pairs, we use a parameter that gives each period ratio room to float. We picked our priors empirically from our own preliminary analysis of where the stable orbits existed. Our prior on the period ratio between b:c and c:d is a uniform distribution between 1.8 and 2.4. For the d:e period ratio, we choose a narrower uniform prior between 1.8 and 2.2, because we found all of the dynamically stable orbits were in this more narrow range and limiting it as such improved the efficiency of finding dynamically stable orbits (Section \ref{sec:dynamics}). We will show in Section \ref{sec:stablecoplanar} that our choices for our priors did not exclude stable orbits. We note that we effectively replaced the parameters for the semi-major axes of the outer planets with their period ratios, so we did not reduce the number of parameters in our MCMC fit even though the parameter space has shrunk. We will refer to this orbital fit as the ``Near 1:2:4:8" fit, which as the naming implies, only places the period ratios near resonance and does not guarantee the planets are indeed in resonance at all.

We initialized the walkers using coplanar solutions, which delayed convergence and caused the walkers to take a considerable amount of time to fully explore all of the allowed parameter space. We ran our parallel-tempered sampler with 15 temperatures and 1500 walkers per temperature for 75,000 steps, after a burn in of 495,000 steps that was chosen using the same metric for convergence as Section \ref{sec:noconstrain}. We performed the same thinning of the chains by a factor of 75. The resulting posterior was taken from the lowest temperature walkers and has 1,500,000 samples. The posteriors are plotted in Figure \ref{fig:posteriors} and reported in Table \ref{table:mcmcfits}. 

\subsection{Near 1:2:4:8 Period Ratio Coplanar Low-$e$ Orbits}\label{sec:1248coplanar}
Lastly, we looked at coplanar resonant orbits. We applied both the coplanarity and period ratio constraints from Sections \ref{sec:coplanar} and \ref{sec:1248}. We also applied an additional constraint that the eccentricity of all of the orbits have to be less than 0.2. In Section \ref{sec:coplanar}, we found that $< 2\%$ of the coplanar orbits have $e_d < 0.2$. From preliminary analysis done concurrently with the orbit fits, we could only find stable orbits when all planets had $e < 0.2$. This fact will be further reinforced by the analysis in Section \ref{sec:stablecoplanar}. Thus, we do not believe we lost stable orbits by applying this additional constraint, and merely improved the efficiency of finding stable orbits. We will refer to this orbital fit as the ``Near 1:2:4:8 Coplanar" fit, which, like the Near 1:2:4:8 fit, does not guarantee the planets are actually in resonance. 

For this 20-parameter orbit fit, we used a parallel-tempered sampler with 15 temperatures and 1500 walkers per temperature. We ran each walker for 125,000 steps, after an initial burn in of 95,000 steps. Convergence was confirmed using the metrics defined in Section \ref{sec:noconstrain}. We again thinned the chains by a factor of 75, and formed our posterior distribution from the lowest temperature chains. This resulted in 2,499,000 samples of the posterior.  The posteriors are plotted in Figure \ref{fig:posteriors} and reported in Table \ref{table:mcmcfits}. 

As we expected, the posterior for the eccentricity of planet d runs up right against our prior bounds. Without stability constraints, higher eccentricity orbits are favored. Just like in the Coplanar orbit fit, we find an orbital inclination for the system in the 20\degr~and 30\degr~range. However, $\Omega$ is now in agreement with that found in \citet{Konopacky2016} for coplanar orbits, unlike our previous orbit fits. It is likely this was a small family of orbits that were not represented in the $1\sigma$ range of our previous analyses. 

\subsection{Goodness of Fit}\label{sec:goodfit}
We used the reduced chi-squared ($\chi^2_{\nu}$) statistic to measure the goodness of fit of a model. Since the highest likelihood model often does not represent the whole posterior of possible orbital configurations, we compute $\chi^2_{\nu}$ on 1000 randomly drawn allowed orbits for each model. We list the 16th, 50th, and 84th percentiles in Table \ref{table:mcmcfits}. For the Unconstrained model, we found $\chi^2_{\nu} \approx 1$, indicating the unconstrained Keplerian orbits can suitably describe the data as one might expect if the uncertainties are estimated accurately, given it is a physical model. 
The other three models have $\chi^2_{\nu}$ similarly close to unity, showing they also fit the data well. 

We also investigated if more-restrictive models with additional, dynamically-motivated constraints better describe the data than the fiducial Unconstrained case. We calculated the Bayesian Information Criteria \citep[BIC;][]{Schwarz1978,Liddle2007} as a simplified alternative to full Bayesian model comparison. The BIC assesses how well a model fits the data and penalizes models that have more free parameters. Models with lower BIC are preferred. We define the $\Delta BIC$ as the difference between the BICs of a more restrictive model and the median BIC of the Unconstrained model. We also calculate $\Delta BIC$ using the same 1000 randomly drawn orbits for each model, and list the 16th, 50th, and 84th percentiles for this value in Table \ref{table:mcmcfits}.

We find the $\Delta BIC$ is negative for the other three models relative to the Unconstrained fit.
This indicates that adding constraints that tend the data towards what we believe are stable orbits makes the fits better, as we discard some parameter space containing likely unstable orbits that do not reflect reality. 
We also note that $\chi^2_{\nu}$ and $\Delta BIC$ are not perfect metrics as they only consider the number of free parameters in the models, and not the total parameter space being considered. In particular, when we limit the period ratios, this does not decrease the number of free parameters while significantly limiting the space of possible orbits. Thus, we see these goodness of fit metrics favor the coplanar solutions as they explicitly reduce the number of parameters in the model. It would be better to have computed the Bayes factor between each pair of models to more rigorously compare models, but the Bayes factor is computationally difficult to calculate with a high-dimensional problem like this and our MCMC samplers were only set up to perform parameter estimation.
Because of this, we do not think it is valid to conclude from solely these two metrics that coplanar orbits are favored. However, we can assert that adding constraints to the orbit fit does not worsen the fit from the fiducial case, and thus the constraints are reasonable given the current astrometric data. This conclusion agrees with the analysis from \citet{Konopacky2016},  who found that coplanar orbits and orbits near 1:2:4:8 period ratios were fully consistent with the Keck astrometry. While the metrics we have employed cannot decide which orbit model should be favored, the stability constraints to the system that are investigated in the following section will clearly show what the realistic orbits are. 

\section{Dynamical Constraints }\label{sec:dynamics}
Keplerian motion is not the only constraint on the orbits of the planets. We also know that these four planets must also have survived from their formation up to this point. The latest estimate for the age of the star is $42^{+6}_{-4}$~Myr old \citep{Bell2015}, based on its membership in the Columba moving group \citep{Torres2008,Zuckerman2011}. This stellar age is further supported by interferometric measurements of the stellar radius \citep{Baines2012}. These orbits must have been stable for roughly the lifetime of the star, since giant planets likely formed quickly before the gas disk dispersed in the first few Myr \citep{Williams2011}. As the gas disk is difficult to model and exists for only a short period of the system's lifetime, we do not simulate the time between planet formation and gas disk dispersal. There could be additional bodies in the system, but it is impractical to consider them without making assumptions on their nature. Instead, our analysis will focus on eliminating unstable orbital configurations based on the four planets alone. If additional bodies are detected, they could further constrain these orbits.

Thus, we investigated which orbital configurations allowed by astrometric measurements are also stable if we simulate the four planets' orbits backwards in time for 40~Myr. In this section, we will apply this dynamical constraint on each of the four orbit fits (Unconstrained, Coplanar, Near 1:2:4:8, Near 1:2:4:8 Coplanar), and investigate the family of stable orbits that arise. 

\subsection{Stability of Orbital Models}\label{sec:stableorbits}
We used the \texttt{REBOUND} $N$-body simulation package \citep{Rein2012} with the WHFast integrator \citep{Rein2015}. To set up a simulation, we added particles for planets e, d, c, b in that order,
using a chosen set of orbital parameters from our fits and placing them at the predicted location on MJD 56609, the date of the first GPI epoch. We drew masses for each of the planets in a process described in the following paragraph, and set the primary mass to be the stellar mass from our orbit fits. We then reversed the present velocities of the planets and integrated the system for 40~Myr to simulate the past dynamical history of the system, using fixed timesteps equal to 1\% of planet e's initial orbital period. We considered a configuration unstable if two planets passed too close, or if one planet was ejected from the system. We considered an encounter too close if any two planets passed with a distance less than the initial mutual Hill radius of planets d and e, which we approximated as
\begin{equation}
    R_{Hd,e} = a_e \left(\frac{M_e + M_d}{3M_\star} \right)^{1/3},
\end{equation}
where $M_e, M_d, M_\star$ refer to the masses of planet e, planet d, and the primary respectively. We considered a planet to be ejected if it moved further than 500 au from the star. Any orbit that survived for 40~Myr without encountering either condition is considered stable. 

To assess the dynamical stability of allowed orbital configurations for each model, we performed rejection sampling to assess which orbit models contained significant amounts of stable orbits. We drew one million random orbital configurations from each of the four model posteriors. The orbit fits do not specify the mass of the planets so we needed to add additional parameters for them. For simplicity, we set planets c, d, and e to be equal in mass as we would expect due to their similar luminosities \citep{Marois2010}. We drew the mass of these planets, $M_{cde}$, from a uniform prior between 4 and 11~$M_{\rm{Jup}}$ to encompass the uncertainty on the luminosity-derived masses from \citet{Marois2010}. We drew the mass of planet b, $M_b$, from a uniform prior between 3 $M_{\rm{Jup}}$ and $M_{cde}$ to account for its lower luminosity. We will discuss using a more informative prior based on the planets' luminosities in Section \ref{sec:masses}. We ran each of the configurations through the \texttt{REBOUND} setup described previously. Our dynamical stability prior sets the probability of an orbit to be 0 if the system is not stable, and 1 if it is stable, discarding the unstable orbits in our rejection sampling. In Table \ref{table:mcmcfits}, we record the number of stable orbits from each of the configurations. 

We found that the Unconstrained and Coplanar orbit solutions did not yield any stable orbits after one million draws. Especially for the Unconstrained case, the lack of stable draws does not mean that these models are inconsistent with stable orbits, but rather that the islands of stability in this high dimensional space are small and were not sampled even after millions of MCMC draws.
Simply, these models do not currently allow for a practical search of stable orbits. 

Both models that assume that the planets' orbital periods are near the 1:2:4:8 period ratio do yield stable orbits, with the model not assuming coplanarity, the Near 1:2:4:8 model, resulting in just one stable orbit after one million draws. This model encompasses all of the parameter space explored by the Near 1:2:4:8 Coplanar model so it is the more general model. We explored this model further by running twenty million \texttt{REBOUND} simulations in total, leading to 50 stable orbits. 

We find that for masses of the inner three planets greater than 5 M$_{Jup}$, the maximum mutual inclination between any pair of planets in a stable system is $< 8^{\circ}$, although we only have few samples in this regime (14 stable orbits spanning 8 unique present-day orbital configurations). We also do not find stable orbits above $6~M_{\rm{Jup}}$, which likely reiterates the difficulty of finding stable noncoplanar orbits due to the high-dimensionality of the problem. Thus, with the limited orbital arcs we have so far, looking for noncoplanar stable orbits is impractical. Since in Section \ref{sec:orbitfit} we found that our astrometry is consistent with the system being coplanar, we will focus on those orbits since we can find many stable orbits with this assumption (hundreds per million tries). We will leave the thorough exploration of orbits with mutual inclinations for future work with longer astrometric baselines and more computation time. However, in our preliminary analysis, it seems that the mutual inclinations are probably small in order for the system to be stable.

\subsection{Stable Coplanar Orbital Solutions}\label{sec:stablecoplanar}
\begin{figure*}[t]
\plotone{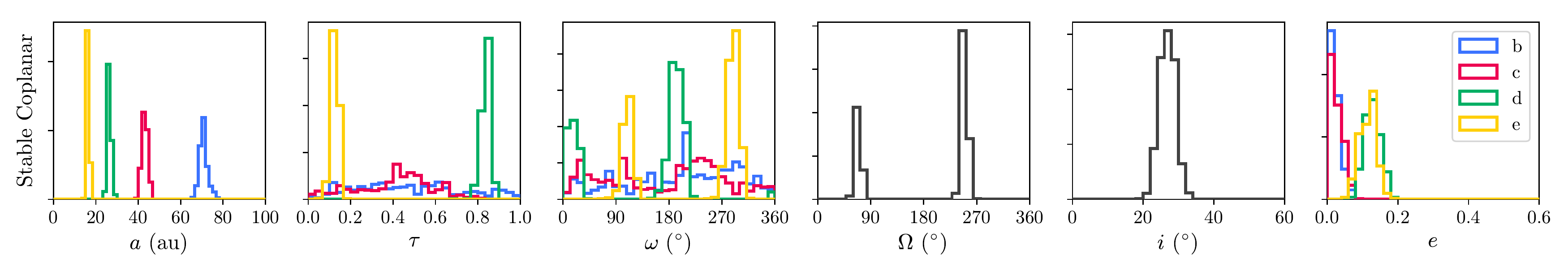}
\caption{Posterior of stable orbital elements for coplanar configurations of the four planets. These posteriors show all stable orbits with $M_{cde} > 4~M_{\rm{Jup}}$ and $3~M_{\rm{Jup}} < M_{b} < M_{cde}$. As discussed in Section \ref{sec:stablecoplanar}, solutions with higher planet masses lie within a smaller region of this space. 
\label{fig:posteriors-stable}}
\end{figure*}

For the rest of the analysis, we focus on the orbital parameters from the Near 1:2:4:8 Coplanar fit. We increase the 
number of $N$-body simulations from one million to twenty-two million, obtaining 9792 stable orbital configurations. We plot the initial osculating orbital elements (i.e., those on MJD 56609) of these stable orbits in Figure \ref{fig:posteriors-stable} and list them in Table \ref{table:stablefits}. 

\begin{deluxetable}{cc|c}
\tablecaption{Stable Coplanar Orbital Parameters of HR 8799 bcde \label{table:stablefits} }
\tablewidth{0pt} 
\tablehead{ 
Body  &  \shortstack[c]{\\ \\Orbital\\Element}  & {\shortstack[c]{Stable\\Coplanar}}  }
\startdata
b & $a_b$ (au)        & $70.8^{+0.19}_{-0.18}$       \\
 & $\tau_b$           & $0.46^{+0.31}_{-0.26}$       \\
 & $\omega_b$ (\degr) & $87 \pm 58$  \\
 & $\Omega_b$ (\degr) & $67.9^{+5.9}_{-5.2}$  \\
 & $i_b$ (\degr)      & $26.8 \pm 2.3$   \\
 & $e_b$              & $0.018^{+0.018}_{-0.013}$  \\
c & $a_c$ (au)        & $43.1^{+1.3}_{-1.4}$  \\
 & $\tau_c$           & $0.43^{+0.15}_{-0.24}$  \\
 & $\omega_c$ (\degr) & $67^{+59}_{-39}$    \\
 & $\Omega_c$ (\degr) & $67.9^{+5.9}_{-5.2}$      \\
 & $i_c$ (\degr)      & $26.8 \pm 2.3$   \\
 & $e_c$              & $0.022^{+0.023}_{-0.017}$  \\
d & $a_d$ (au)        & $26.2^{+0.9}_{-0.7}$  \\
 & $\tau_d$           & $0.839^{+0.020}_{-0.017}$  \\
 & $\omega_d$ (\degr) & $17^{+12}_{-11}$      \\
 & $\Omega_d$ (\degr) & $67.9^{+5.9}_{-5.2}$      \\
 & $i_d$ (\degr)      & $26.8 \pm 2.3$   \\
 & $e_d$              & $0.129^{+0.022}_{-0.025}$  \\
e & $a_e$ (au)        & $16.2 \pm 0.5$  \\
 & $\tau_e$           & $0.124^{+0.019}_{-0.013}$  \\
 & $\omega_e$ (\degr) & $110 \pm 9$      \\
 & $\Omega_e$ (\degr) & $67.9^{+5.9}_{-5.2}$      \\
 & $i_e$ (\degr)      & $26.8 \pm 2.3$   \\
 & $e_e$              & $0.118^{+0.019}_{-0.028}$  \\
A & Parallax (mas)& $24.30^{+0.49}_{-0.69}$ \\
 & $M_{\star}$ ($M_\odot$)& $1.47^{+0.11}_{-0.08}$  \\
 \hline
 & $\chi^2_{\nu}$ & $1.01^{+0.06}_{-0.05}$ \\
 & $\Delta BIC$ & $-22^{+6}_{-4}$ \\  
\enddata
\tablecomments{Values are reported in the same way as Table \ref{table:mcmcfits}}
\end{deluxetable}

\begin{figure*}[t]
\plotone{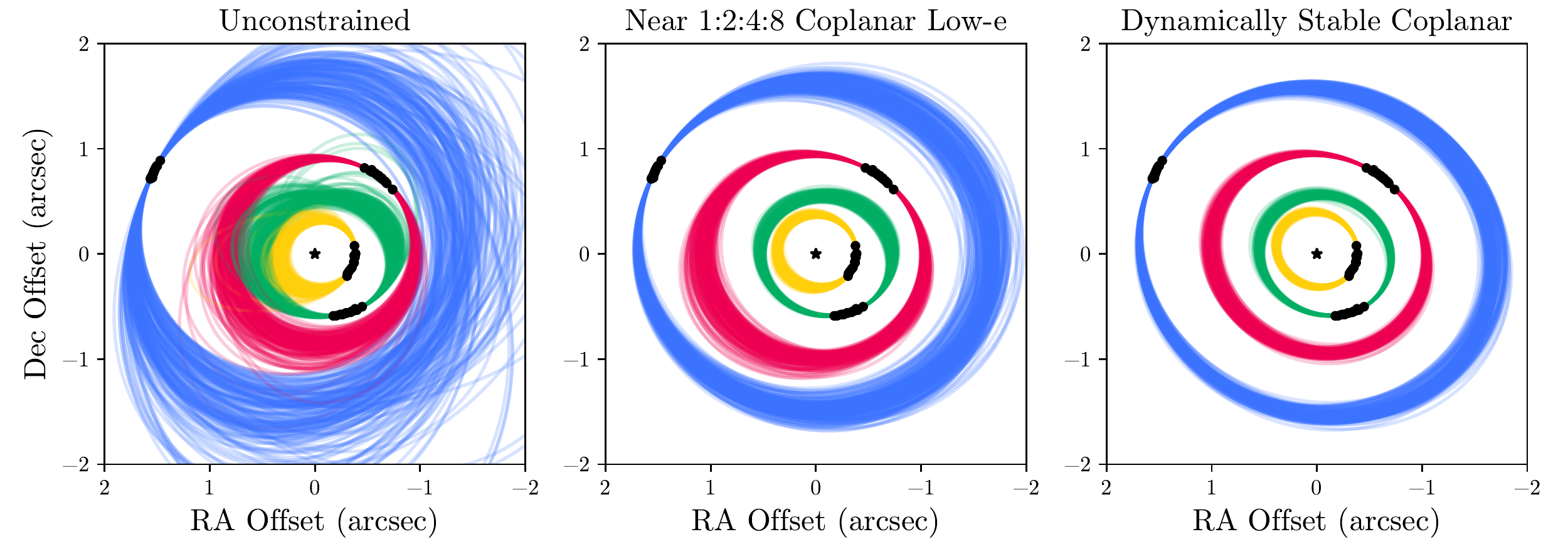}
\caption{A comparison of 200 allowed orbits from the Unconstrained (Section \ref{sec:noconstrain}), Near 1:2:4:8 Coplanar (Section \ref{sec:1248coplanar}), and dynamically stable coplanar solutions (Section \ref{sec:stablecoplanar}) projected onto the sky plane. The black star in the middle represents the location of the star, the black circles are the measured astrometry (uncertainties too small to show on this scale), and the current orbit for each planet is colored in the same way as Figures \ref{fig:posteriors} and \ref{fig:posteriors-stable} (i.e., planet b is blue, c is red, d is green, and e is yellow). 
\label{fig:sky-proj}}
\end{figure*}

We find that the posteriors have tightened significantly after applying the dynamical stability constraint. Figure \ref{fig:sky-proj} visually compares the spread of possible orbits on the 2-D sky plane for the orbit fits with increasing constraints placed on them. The stable coplanar orbits appear as a well defined ellipse with minimal uncertainty for each planet's orbit. This is also reflected visually and numerically in the posterior percentiles. The middle 68\%, the difference between the 84th and 16th percentiles, of the semi-major axes of the planets decreased by 1.5 to 4.5 times when compared to the Unconstrained case, and by a factor of 1.17 to 1.50 when compared to the Near 1:2:4:8 Coplanar fit that the stable orbits were drawn from. Similarly, the middle 68\% of the eccentricities also decreased by a factor between 2.2 and 4.7 compared to the Unconstrained fits. In fact, the fractional uncertainty on the semi-major axes is about the same as the fractional uncertainty of the Gaia DR1 parallax of the system ($\approx$2\%). The inclusion of the parallax from Gaia Data Release 2, released after this analysis was completed, should reduce its contribution to the semi-major axis and total system mass uncertainties by a factor of 7 \citep{Gaia2018}. We have chosen not to rerun our analysis since the conclusions in this paper do not strongly depend on the exact semi-major axes of the orbits, and we will leave this for a future work.

The stable orbits, despite being much more restrictive, are good fits to the data. Aggregating 1000 random orbits, we find a $\chi^2_{\nu} = 1.01^{+0.06}_{-0.05}$ that is just as good as the fiducial Unconstrained model. The $\Delta BIC$ is similarly comparable to the other models. Although, once again we note that BIC does not account for the narrower parameter space due to the additional stability constraint. We conclude that these stable orbits are a small, but allowed part of a much larger space that we have explored through our Bayesian analysis. 

\begin{figure}[t]
\plotone{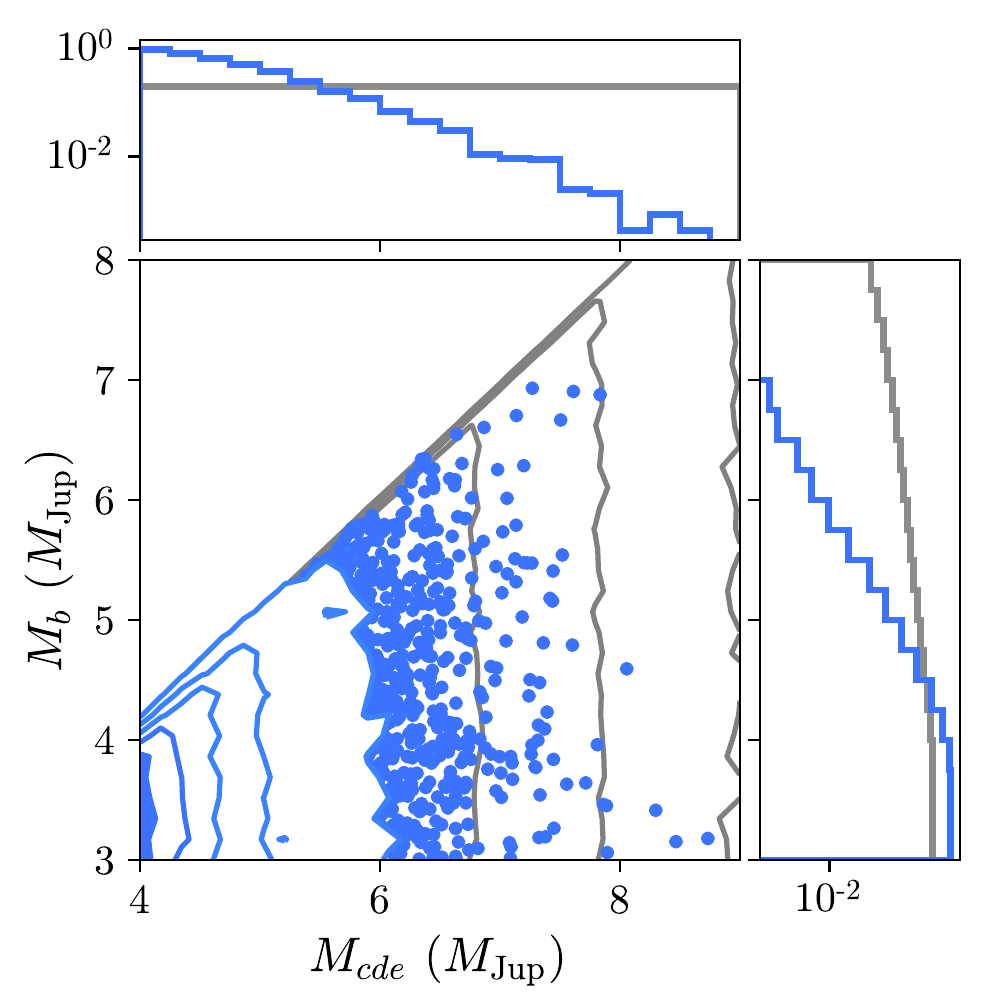}
\caption{The distribution of masses of the stable orbits from Section \ref{sec:stablecoplanar} (blue) and comparison to the priors from which the masses were drawn (gray). The main plot in the bottom left shows the 2-D distribution of masses. The contour lines represent 15th, 35th, 55th, 75th, and 95th percentiles of the distribution, with everything outside the 95th percentile plotted individually as points. The top and right panels show 1-D histograms for $M_{cde}$ and $M_{b}$ respectively, with the frequency in each bin plotted on a logarithmic scale to highlight the high mass bins. The gray priors are plotted in the same fashion as the blue posteriors. 
\label{fig:mass-hist}}
\end{figure}

The masses of the stable configurations are plotted in Figure \ref{fig:mass-hist}. We will discuss mass constraints in Section \ref{sec:masses} in detail. Briefly here, we can see that stable orbits exist with the mass of the inner three planets at almost 9~$M_{\rm{Jup}}$, and separately with the mass of planet b to be nearly 7~$M_{\rm{Jup}}$. Also, the majority of stable orbits we found are low mass. 95.6\% of the orbits have $M_{cde} < 6~M_{\rm{Jup}}$ and 73.0\% of the orbits have $M_{cde} < 5~M_{\rm{Jup}}$. This highlights the difficulty in finding stable high mass solutions when starting with our current orbit fits. 

\begin{figure}[t]
\plotone{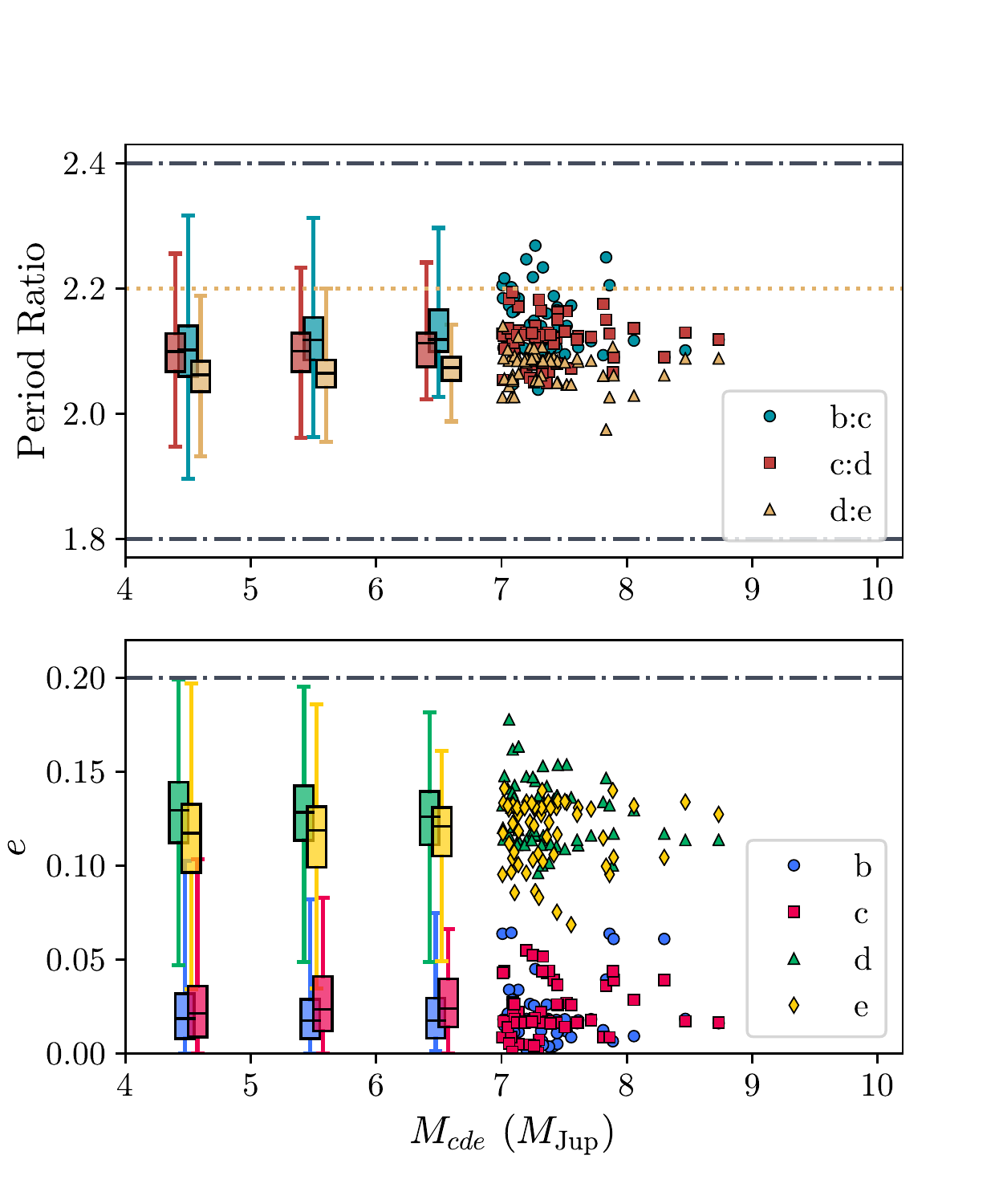}
\caption{The distribution of current period ratios and eccentricities as a function of mass of the inner three planets for stable orbits. For $M_{cde} < 7~M_{\rm{Jup}}$, the data is binned into one box plot per $M_{\rm{Jup}}$. Each box shows the 25th, 50th, and 75th percentiles of the given distribution, while the whiskers show the extrema. Above 7~$M_{\rm{Jup}}$, points are plotted individually as they are sparse enough. The bounds of the priors are plotted as gray dot-dashed lines, except for the upper bound of the d:e period ratio which is the yellow dotted line. These plots show how the range of allowed period ratios and eccentricities decrease as planet mass increases. Above $M_{cde} > 6~M_{\rm{Jup}}$, the full range of stable orbits are not near the prior bounds. 
\label{fig:pr-e-mass}}
\end{figure}

With these stable orbits, we can look at the mass dependence on the orbital parameters to justify our choices of prior constraints in doing the Near 1:2:4:8 Coplanar orbit fit. In Figure \ref{fig:pr-e-mass}, we plot the range of period ratios and eccentricities of stable orbits as a function of $M_{cde}$. We see that for $M_{cde} > 6~M_{\rm{Jup}}$, none of the period ratios or the eccentricities are close to the bounds set by our priors. Below 6~$M_{\rm{Jup}}$, the period ratio of planet d to e as well as the eccentricities of d and e are near the upper bound in the extreme case, indicating our prior may be excluding some low-mass stable orbits. Since the interquartile range of these parameters is far away from these bounds, only a few extreme low-mass cases have been excluded, so the effect should be minimal.
As the masses increase, we see the range in the allowed parameter space decreases, indicating that the highest mass stable orbits reside in a subspace of the parameters we are exploring. Thus, we conclude that we are not unnecessarily excluding stable orbit configurations with our choice of priors that were designed to improve the efficiency of finding stable orbits. 

There are several notable features in our posteriors of stable orbital configurations. The bimodality of the eccentricity posteriors is clear. The outer planets b and c have $e \sim 0 $ while the inner planets d and e have $e \sim 0.1$. These eccentricities agree well with what was found by \citet{Gozd2014} who migrated planets into resonance lock, rotated the orientations to match the astrometry, and selected orbital configurations with a $\chi^2_{\nu}$ cutoff. Given that this conclusion was reached by two completely different analysis methods, the fact the inner two planets have slightly eccentric orbits while the outer two planets are in near-circular orbits is a notable result that seems to be required for most stable orbital configurations that are consistent with the measured astrometry. 
The increased eccentricities of planets d and e, and the proximity of all four planets to 1:2:4:8 period commensurabilities, are consistent with
an early evolutionary period of
convergent inward migration of all four planets, trapping
of planet pairs d \& e and c \& d into 2:1 resonances, and pumping of the orbital eccentricities of d and e by continued migration while in resonance
lock (e.g., \citealt{Yu2001}; see also section \ref{sec:resonances}). 

Also, comparing our stable orbits with those of \citet{Gozd2014} and \citet{Gozd2018}, we note that most orbital parameters agree fairly well except for the semi-major axes of the planets, which we find to be significantly larger. For example, only 0.11\% of our orbital solutions have $a_c \leq 39.4$ au, the best fit solution of \citet{Gozd2014}. Our uncertainties in parallax and stellar mass are consistent with the fixed parallax and stellar mass they used, so the difference in $a$ is not just a result of different system parameters. We also generally have larger uncertainties on our values, which can be due to a combination of allowing lower mass orbits, not strictly enforcing 1:2:4:8 resonance lock, and a more systematic exploration of parameter space. 

As the orbits are coplanar, each planet's argument of periastron can be used to measure the relative orientation of the planets' orbits. As both planet b and c have extremely low eccentricities, $\omega$ is basically unconstrained for these planets and is not notable since their orbits are near circular. The significant nonzero eccentricity of planets d and e however correspond to sharp peaks in $\omega$ for both planets. While a broader peak was already seen in $\omega_d$ in the Near 1:2:4:8 Coplanar fits before enforcing stability constraints, the addition of dynamical stability has disallowed circular orbits of planet e, giving rise to a sharp peak in $\omega_e$. Interestingly, the orientation of the orbits of planets d and e are not aligned, with $\omega_e - \omega_d = 94_{-9}^{+11}$ degrees, essentially perpendicular to being aligned. We note that \citet{Gozd2014} also found a similar result. 

The period ratios of the planets shown in Figure \ref{fig:pr-e-mass} heavily favor period ratios above the nominal 2:1. We find period ratios $P_b/P_c = 2.11 \pm 0.06$, $P_c/P_d = 2.10^{+0.04}_{-0.05}$, and $P_d/P_e = 2.06^{+0.03}_{-0.04}$. For $P_c/P_d$ and $P_d/P_e$, the data favors period ratios above two. This can been seen by computing the period ratios using the median $a$ for the Near 1:2:4:8 Coplanar fits from Table \ref{table:mcmcfits} that do not have a dynamical prior applied. It seems that these period ratios are at these high values to satisfy the astrometry. 
For $P_b/P_c$, the data allows both period ratios above and below two, so having it strongly favor values above 2 (only 2\% of stable configurations have $P_b/P_c < 2$) indicates that spacing the two planets slightly further apart enhances stability. The period ratios driven by the astrometry could be indicative of a primordial period ratio. In particular, the planets could have experienced eccentricity damping while in resonance and were repelled to period ratios greater than 2 while still maintaining resonance lock \citep{Lithwick2012,Batygin2013}. As disk gas is the primary mechanism for eccentricity dissipation of Jupiter-mass planets at large-separations,\footnote{Note, however, that
damping of planetary eccentricity by gas dynamical friction
does not conserve the planet's orbital angular momentum,
contrary to damping of eccentricity by tidal
dissipation in the planet; the latter, not the former, is 
considered by \citet{Lithwick2012} and \citet{Batygin2013}.
However, the general mechanism of resonant repulsion also occurs for eccentricity damping by gas dynamical friction (e.g., as simulated for giant planets at wide sparations by \citealt{Dong2016}).
} this may indicate that the planets were in or near their current location during the gas disk stage \citep{Dong2016}.

We found the system has an inclination of $i = 26\fdg8 \pm 2\fdg3$ and longitude of ascending node of $\Omega = 68\fdg0^{+5.9}_{-5.3}$, consistent with the work by \citet{Konopacky2016} fitting coplanar orbits, but a few times more precise. Both $i$ and $\Omega$ match the debris disk inclination of $26^{\circ} \pm 3^{\circ}$ and position angle of $62^{\circ} \pm 3^{\circ}$ derived from far-infrared \textit{Herschel} observations \citep{Matthews2014}.
While the inclination is also consistent with the millimeter observations of the debris disk by the Submillimeter Array (SMA) and the Atacama Large Millimeter/submillimeter Array (ALMA), $\Omega$ is
higher than the position angle of the disk of $35\fdg6^{+9.4}_{-10.1}$ in the millimeter \citep{Wilner2018}.
This implies the disk in the millimeter is mutually inclined from the planetary orbital plane by $16^{+22}_{-11}$ degrees.
If this offset is real, we would be observing a process that decouples the millimeter planetesimals from the planets and smaller dust grains probed by \textit{Herschel}. However, it is not clear what could cause that. As reported in Section \ref{sec:stableorbits}, we do not find high mass solutions that are mutually inclined by over 8\degr, so it seems unlikely that planet b is torquing the disk, although we cannot definitely exclude planet b being inclined from the rest of the planets. 
Still, it would not explain why the smaller dust seen by \textit{Herschel} are indeed coplanar with the planets. Deeper observations of the debris disk are needed to determine if the debris disk is coplanar to the planets. 

Previous works have considered the need for an additional planet to carve the inner edge of the outer belt of debris \citep{Booth2016,Read2018}. Here we investigated whether planet b is consistent with sculpting the inner edge of the outer belt, assuming the planets are coplanar or nearly so to the disk; the possible millimeter-wave offset
of $16^{+22}_{-11}$ deg cited above is assumed negligible in this regard.
Since planet b's orbit is likely near-circular, with 95\% of the allowed stable orbits having $e_b < 0.05$, we can compute the clearing zone of planet b using the following equation from \citet{Morrison2015} for the outer edge of a planet's chaotic zone, validated for high-mass planets like HR 8799 b:
\begin{equation}
    R_{in} = a_p + 1.7 a_p (M_p/M_\star)^{0.31}.
\end{equation}
Here $R_{in}$ is the inner radius of the disk and corresponds to the outer edge of a planet's clearing zone, $a_p$ is the semi-major axis of the planet, $M_p$ is the mass of the planet, and $M_\star$ is the mass of the star. Plugging in the numbers from our dynamically stable orbits, we find $R_{in} = 89^{+3}_{-2}$~au when considering all stable orbits, and $R_{in} = 93^{+3}_{-2}$~au when considering only stable orbits with $M_b > 5$~$M_{\rm{Jup}}$. When compared to the inner edge of $104^{+8}_{-12}$~au derived by \citet{Wilner2018}, our median value of the inner edge when only considering $M_b > 5~M_{\rm{Jup}}$ is consistent with their middle 68\% credible interval, while our median value when considering all of our stable solutions equally is slightly below this credible interval. However, given that our quoted numbers on the inner disk edge depends on our priors on the mass of the planet and given the uncertainty in the inner disk edge \citep{Booth2016,Wilner2018}, a $1\sigma$ disagreement is not significant. Our orbit fits place planet b at a location consistent with sculpting the inner edge of the debris disk, although finer studies of the dynamical interactions of system and more refined system parameters will help clarify the picture. 

\subsection{Orbital Resonances}\label{sec:resonances}
Having stable orbits near integer period ratios does not guarantee resonance. To explore possible orbital resonances in our stable configurations, we saved the state of each stable system every 200 years using the \textit{SimulationArchive} feature of \texttt{REBOUND} \citep{Rein2017}. We looked at the resonant angles of each system as a function of time to infer the resonant nature of the system: planets in resonance will have a corresponding resonant angle that is librating, but the angle will circulate if the planets are not in resonance. For this four planet system, we looked at nine resonant angles. The first six angles are two-body resonant angles that look at whether consecutive planets are in 2:1 resonance. These 2:1 resonant angles are defined by
\begin{equation} \label{eq:21res}
    \theta_{i:j,k} = \lambda_j - 2 \lambda_i + \varpi_k, \quad k \in \{i,j\}.
\end{equation}
Here, $\varpi = \Omega + \omega$ is the longitude of periastron, and $\lambda = \varpi + M$ is the mean longitude where $M$ is the mean anomaly. The labels $i$ and $j$ refer to the labels of a consecutive pair of planets with the inner planet being $j$, and $k$ refers to either $i$ or $j$, resulting in two resonant angles per pair of planets and thus six two-body resonant angles in total for the three consecutive pairs of planets. For example, the 2:1 resonant angle for planets c and d using planet d's $\varpi$ would be written as $\theta_{c:d,d}$.
In a similar notation, the three-body 1:2:4 Laplace resonance can be written as 
\begin{equation} \label{eq:421res}
    \theta_{i:j:k} = \lambda_k - 3 \lambda_j + 2 \lambda_i,
\end{equation}
where the innermost planet is $k$, and the outermost planet is $i$. Lastly, we looked at the same four-body 1:2:4:8 resonant angle as \citet{Gozd2014}:
\begin{equation} \label{eq:8421res}
    \theta_{b:c:d:e} = \lambda_e - 2\lambda_d - \lambda_c + 2\lambda_b.
\end{equation}

\begin{figure}
\epsscale{1.1}
\plotone{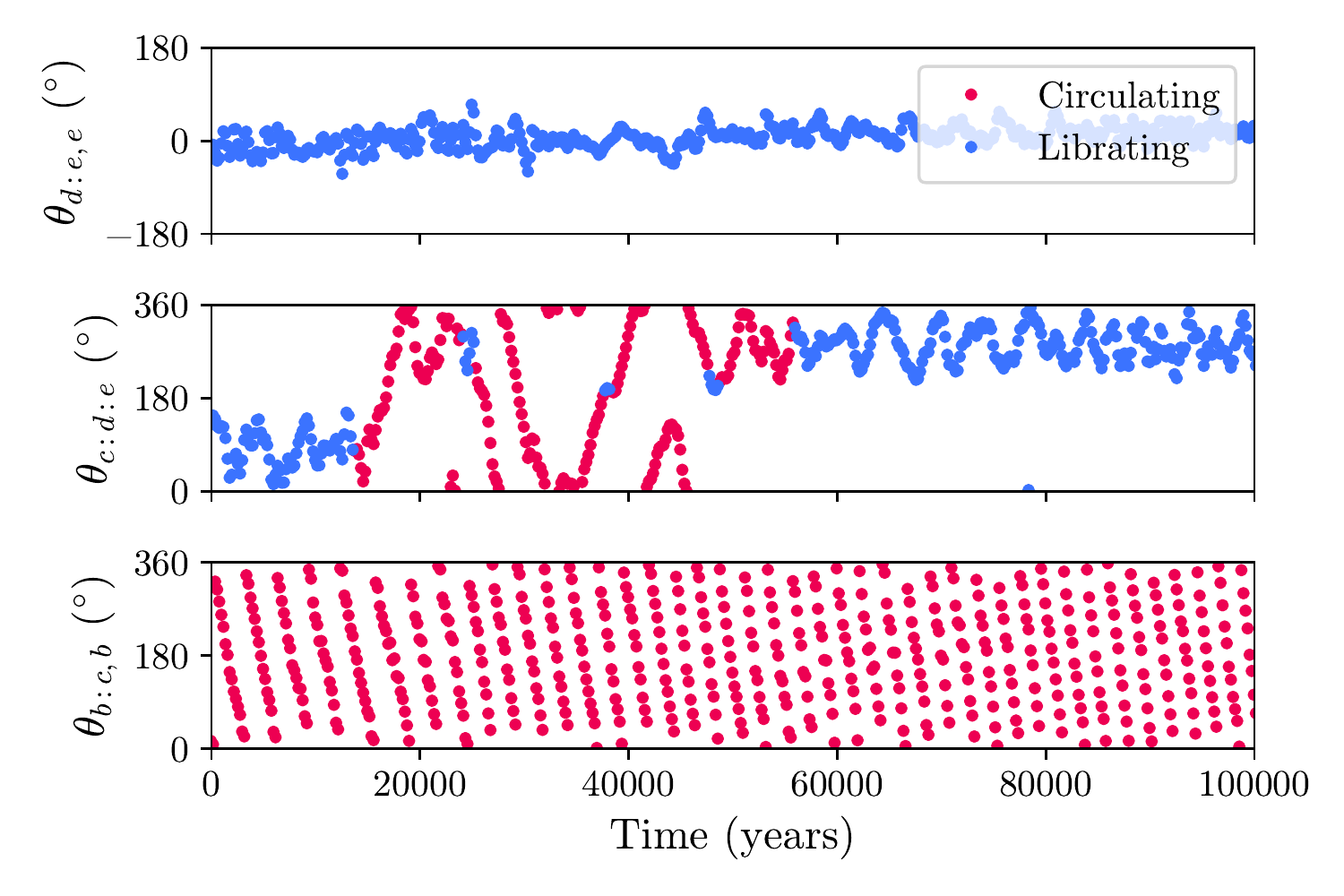}
\caption{Demonstration of the algorithm to identify librating and circulating segments of each critical angle. The top, middle, and bottom plots show a purely librating, transitioning, and purely circulating angle respectively. The points colored red show regions identified as circulating, and the points colored blue show regions identified as librating. In this example, $\theta_{d:e,e}$ is always librating with a librating center of 0\degr~and a libration amplitude of 36\degr. $\theta_{c:d:e}$ is librating only 33\% of the time and $\theta_{b:c,b}$ is circulating 98\% of the time so librating amplitudes and centers are not well defined. 
\label{fig:lib_algo_demo}}
\end{figure}

\begin{figure*}
\plotone{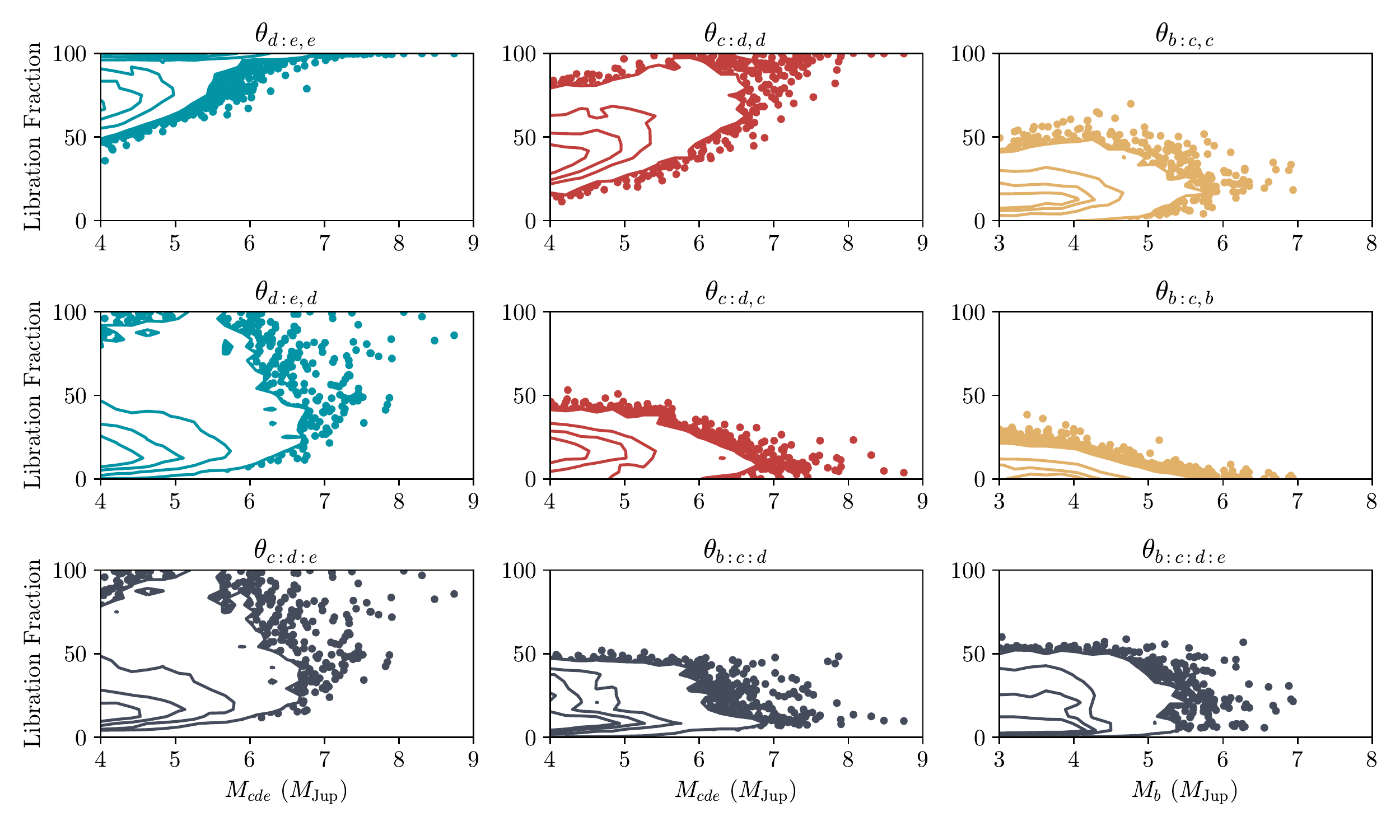}
\caption{Distribution of systems in the space of libration fraction versus planet mass. Contours are 25th, 50th, 75th, and 97th percentiles (e.g., 97\% of systems lie within the 97th percentile contour). Above that, individual points that correspond to particular stable orbital configurations are plotted. Libration fractions of 100\% indicate resonance lock, while libration fractions $<5$\% indicate the planets are probably never in resonance. 
\label{fig:libfrac}}
\end{figure*}

In our simulations, we found that these resonant angles varied in behavior, with some continuously librating (i.e., locked in resonance for 40~Myr), some continuously circulating, and some transitioning between the two over the 40~Myr orbit integration. To analyze all of the simulations uniformly, we developed an algorithm to identify libration and compute the fraction of time a resonant angle is librating or circulating over the course of a simulation. The algorithm takes advantage of the fact that librations oscillate around a fixed value while circulating angles are monotonically changing. Briefly, the algorithm uses a Fourier transform to identify the periodicity of the data, smooths it on that scale, and computes the time derivative of the smoothed angle over the time series. Any sections of the time derivative with significant deviations from zero are deemed circulating and the rest are deemed librating. Figure \ref{fig:libfrac} shows an example of this algorithm classifying librating and circulating sections of a few resonant angles. We note that this method is not perfect, and requires a subjective threshold to determine when a deviation is significant. However, inspecting the results from several resonant angles by eye, the algorithm seems comparable to by-eye identification, and seems to accurately identify resonant angles that are librating continuously (i.e., librating 100\% of the time). For systems which by eye are transitioning quickly or which are always circulating, we estimate that we misidentified $\sim$5\% of the time-series. This error is small, and the gain in having an automated algorithm to uniformly analyze all of these time-series is large. 

We apply this algorithm to all nine resonant angles for each of our simulations. We plot the libration fraction, the fraction of time in the last 40~Myr during which that angle is librating, for each angle in Figure \ref{fig:libfrac}. At low masses ($M_{cde} < 6 M_{\rm{Jup}}$), the scatter in the libration fraction is high for all angles, indicating resonance lock for any subset of the planets is not necessary for stable orbits at low masses. Several of the angles never reach 100\% libration fraction at any mass, indicating that all of the stable orbital configurations we found do not have all four planets in resonant lock. While we do not see any four planet resonant chains, it might be possible they reside in a small island of parameter space that our MCMC did not sample. All resonance angles involving planet b never reach 100\% libration fraction. Planet b may occasionally come into resonance with the inner planets, but does not remain there. 
It is not too surprising that planet b 
does not favor resonance, insofar as the
magnitude of the
resonant potential associated with $\theta_{b:j,b}$
is proportional to the planet's orbital eccentricity
$e_b$ \citep[e.g.,][]{Murray1999}, and its median eccentricity
is the lowest among the four planets.

The inner three planets do favor resonance more than planet b. Above $\sim$6~$M_{\rm{Jup}}$, a large majority of stable orbits have $\theta_{d:e,e}$ librating 100\% of the time. Similarly, above $\sim$7~$M_{\rm{Jup}}$, $\theta_{c:d,d}$ is always librating for most stable orbits. For all masses, these two angles are always librating some of the time. However, the other two resonant angles, $\theta_{d:e,d}$ and $\theta_{c:d,c}$, appear primarily transitioning between libration and circulation, with $\theta_{d:e,d}$ trending to libration at high masses. This behavior is also reflected in the three-body resonant angle between the inner three planets, with only 1.2\% of the stable orbits having this three-body angle librating for at least 90\% of the time. Still, this behavior indicates that the inner planets being in a 1:2:4 three-body resonance is both consistent with the data and dynamically stable for masses up to 8~$M_{\rm{Jup}}$. In these cases, the libration center of $\theta_{c:d:e}$ often jumps between $\sim90\degr$ and $\sim-90\degr$, but typically keeping $|\varpi_d - \varpi_e| \sim 90\degr$. However, the 1:2:4 three-body resonance is not required for stability, even at high masses. 

When the three planets are not locked in resonance together, pairs of planets can be in resonance. We find, averaged across the ensemble of simulations, these two-body resonant angles librate around 0\degr. However, in a single simulation, the libration center can be offset from 0\degr, a phenomenon know as asymmetric libration (e.g., \citealt{Murray-Clay2005}) that is observed for two body angles in other resonant chain systems (e.g., Kepler-80, \citealt{MacDonald2016}) and is caused by the gravitational effect of a third planet.
For the situation where $\theta_{d:e,e}$ librates but $\theta_{d:e,d}$ does not, conjunction of planets d and e always occurs at the periastron of planet e's orbit, but is completely uncorrelated with planet d's orbit. In the case where both $\theta_{d:e,e}$ and $\theta_{c:d,d}$ librate but the three-body angle $\theta_{c:d:e}$ does not, planet e's orbit orients itself so that it is lined up to the conjunction with planet d, while planet d's longitude of periastron is driven by the conjunction with planet c, which is not locked in with planet e. In this case, consecutive planet pairs appear to be locked in resonance for 40~Myr, but three-body resonance lock does not exist.

Having planets d and e and planets c and d locked in two-body resonances fits well with the picture that they were locked in resonance quickly after formation, before the gaseous protoplanetary disk disappeared. After the planets migrated into resonance lock, the eccentricities of e and d were both amplified by their resonant migration in the gas disk and damped by the gas, pushing the planets to larger period ratios. After the gas dispersed, the planets maintained their primordial eccentricities and period ratios, with these parameters only oscillating as the planets exchange energy and angular momentum in resonance.

\subsection{Dynamical Mass Limits }\label{sec:masses}
As we only have a short orbital arc of data, we are limited on the mass constraints we can place based solely on dynamical considerations. We have not yet measured the perturbations of the planets' orbits by each other. Without seeing a significant effect, we cannot place a lower bound on the masses of the planets dynamically. Impractically, we may need to measure the change in the orbital elements of the planets after many orbits, akin to the masses derived from transit timing variations \citep{Agol2005,Holman2005}. Thus, dynamical constraints based on short orbital arcs cannot fully constrain the masses alone. 

Looking at Figure \ref{fig:mass-hist} again, our stable orbits heavily favor low masses, since we cannot place a lower bound on the masses and we used a uniform prior. For high masses, $M_{cde} > 8~M_{\rm{Jup}}$, we have only a few stable orbits, all with the mass of the planet b near the lower-bound of what we would expect, indicating a possible upper limit to the masses of the planets. However, we cannot verify this is because of a lack of sampling of stable orbits. That is, the probability of drawing a stable orbital configuration with $M_{cde} > 8~M_{\rm{Jup}}$ and $M_{b} > 5~M_{\rm{Jup}}$ might be so small that we do not expect to find one with our current sampling. We are potentially limited by the fact that 2.5 million samples of the posterior are not sufficient when the posterior has 20 dimensions and the islands of stability at extremely high masses are extremely small. Thus, there is no indication of a sharp drop-off that would point to a firm upper limit on the mass based on dynamical considerations. In fact, \citet{Gozd2014} and \citet{Gozd2018} found stable orbits above 9~$M_{\rm{Jup}}$ for the inner three planets, indicating higher-mass stable orbits exist if one forces the entire system to be in a 4-planet resonance lock. It might be that we missed those systems since we did not enforce such a global lock and therefore had a larger parameter space of allowed stable orbits. 
If we acknowledge that we are dependent on our choice of priors for the mass of the planets, we can say that 99.9\% of the orbits that are dynamically stable for the last 40~Myr have $M_{cde} < 7.6~M_{\rm{Jup}}$ and $M_{b} < 6.3~M_{\rm{Jup}}$ for this particular choice of prior. These mass upper limits are consistent with the luminosity derived masses of $M_{cde} = 7~M_{\rm{Jup}}$ and $M_{b} = 5~M_{\rm{Jup}}$ based on hot-start evolutionary models \citep{Marois2008,Marois2010}. 

\begin{figure}
\plotone{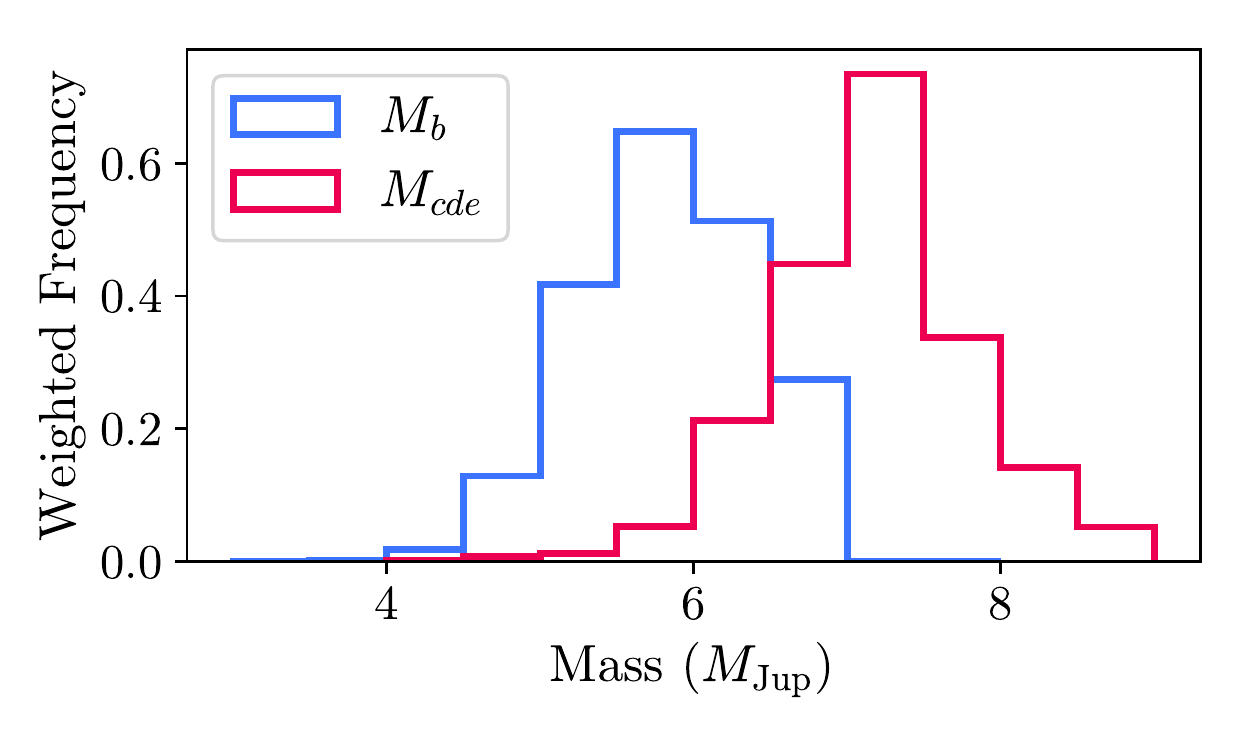}
\caption{The histogram of dynamically stable masses after being weighted by the luminosity prior on the planet masses assuming hot-start evolutionary tracks. Planet b is in blue, and planets c, d, and e are assumed to follow the same red histogram. 
\label{fig:lum_mass}}
\end{figure}

We also investigated using a mass prior based on the measured luminosity and hot-start evolutionary models. Since it is computationally intensive to rerun all 22 million $N$-body simulations, we instead weighted each sample drawn from the previous prior distribution with the relative change in probability due to switching to a luminosity-based prior, resulting in a down-weighting of the lowest mass configurations which are disfavored by the measured luminosities. To accomplish this, we reevaluated the luminosity-derived masses with newest age estimates for the Columba moving group from \citet{Bell2015} which we approximate as a Gaussian distribution of $42 \pm 5$~Myr. This age has better stated uncertainties than the 30-60~Myr range given in \citet{Marois2010}. We still used the same luminosities measured by \citet{Marois2008} and \citet{Marois2010}, as there has not been an update to them with stated uncertainties. Using the \citet{Baraffe2003} hot-start cooling tracks, we get model-dependent masses of $6.0\pm0.7$, $8.7\pm1.0$, $8.7\pm1.0$, and $8.7\pm1.7$~$M_{\rm{Jup}}$ for planets b, c, d, and e respectively. To stay self-consistent with our simulations that fix the masses of the inner three planets to be the same, we choose to use the $8.7\pm1.0~M_{\rm{Jup}}$ of planets c and d for the mass of the planet e also. Otherwise, planet e will tend toward lower masses, and be inconsistent with our simulation assumption of equal masses for the inner three planets. 
Indeed, spectrophotometric measurements from the latest generation of high-contrast imagers confirm that planet e has similar near-infrared fluxes to planet c and d, and not planet b \citep{Zurlo2016,Greenbaum2018}, so our assumption should be robust. Altogether, when combining the luminosity and dynamical constraints for the mass assuming hot-start evolutionary tracks, we get a mass of $5.8 \pm 0.5~M_{\rm{Jup}}$ for planet b and $7.2_{-0.7}^{+0.6}~M_{\rm{Jup}}$ for planets c, d, and e. These mass distributions are plotted in Figure \ref{fig:lum_mass}. Note that our mass estimates depend on the resonances we found the planets to be in, and that higher masses can be achieved by assuming four planet resonance lock \citep{Gozd2014,Gozd2018}.

It is uncertain exactly how bright planets are during the first 100~Myr as it depends on uncertain formation mechanisms. Planet cooling tracks are instead parameterized by a quantity like the initial entropy of the material that formed the planet \citep{Spiegel2012}. Since the hot-start models really are the high-entropy upper limit with regards to planet formation models, the masses estimated assuming these tracks are the lowest masses for the planets. Thus, we can use the hot-start model to quote a lower limit on the mass. Combining the dynamical constraints with the luminosity prior on the masses from the hot-start model, we find 95\% of the stable systems have $M_{b} > 4.9~M_{\rm{Jup}}$ and $M_{cde} > 6.1~M_{\rm{Jup}}$. Alternatively, we can use the upper limits on the masses from dynamical stability alone to constrain the initial conditions of the cooling tracks. Using the same stated luminosities and age of the system and using the \citet{Spiegel2012} warm-start models, our upper limits of $M_{cde} < 7.6~M_{\rm{Jup}}$ and $M_{b} < 6.3~M_{\rm{Jup}}$ would then correspond to lower limits on the initial entropy of $9.5~k_B$ per baryon for the inner three planets and $9.2~k_B$ per baryon for planet b. This excludes the most-extreme cold-start formation models, but is consistent with a range of higher entropy models, as warm- and hot-start models have similar luminosities at this age.

\subsection{Long-Term Dynamical Stability}\label{sec:future}

\begin{figure}
\plotone{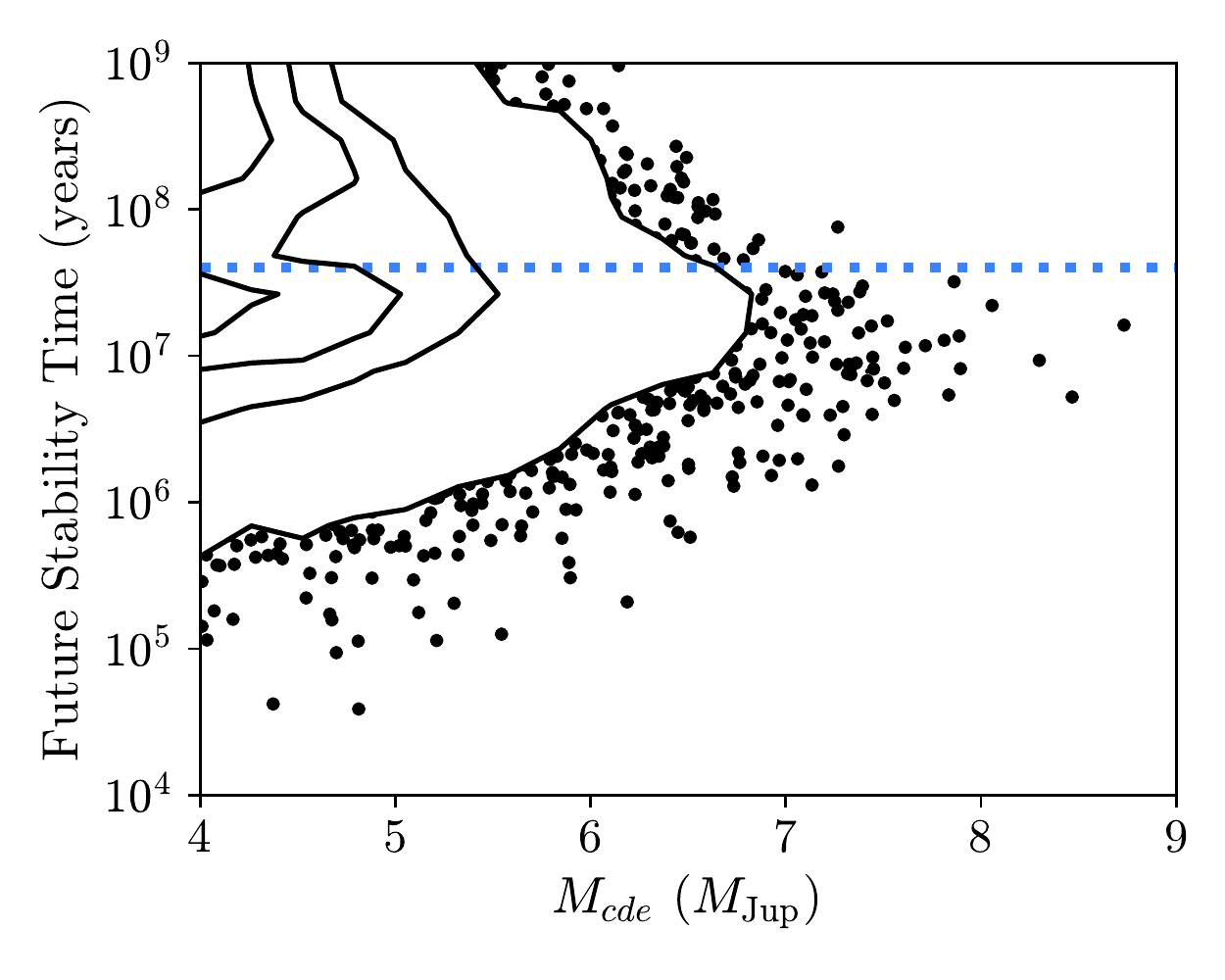}
\caption{The stability time of the dynamically stable orbits integrating forwards in time for up to 1~Gyr. Contours are 25th, 50th, 75th, and 97th percentiles. The horizontal blue dotted line indicates 40~Myr, the time we integrated the systems backwards as reference. 
\label{fig:future}}
\end{figure}

Even though the system has been stable for $\sim$40~Myr, we investigate whether the system we see today is reflective of the final state of system, or whether this configuration is a transient one. We used the saved \textit{SimulationArchive} of each stable orbit, reversed the velocities again, and now integrated them forwards in time for up to 1~Gyr. We used the same criteria to assess stability as before. We plot the amount of time in the future each system is stable as a function of the mass of the inner three planets in Figure \ref{fig:future}. 

We find orbits that are stable for 1~Gyr up to 6~$M_{\rm{Jup}}$. Above 7~$M_{\rm{Jup}}$, none of our orbits are stable for over 100~Myr. Thus, if these planets have masses above 7~$M_{\rm{Jup}}$ as is favored by our combined dynamical and luminosity constraint, then the system is not likely stable. As there is evidence in our own Solar System of dynamical upheaval of our less-tightly-packed gas giants early on \citep{Tsiganis2005}, it is not surprising to find that the HR 8799 system will become unstable. 

Certain resonances do seem to improve stability. In particular, in systems where the three-body angle $\theta_{c:d:e}$ is librating at least 50\% of the time, systems are 10 times less likely to go unstable in 1~Myr and 4 times less likely to become unstable in the 10~Myr than systems with $\theta_{c:d:e}$ librating less than 50\% of the time. There does not appear to be strong correlation between the two-body angles and stability except for $\theta_{b:c,c}$. Systems for which $\theta_{b:c,c}$ is libration more than 30\% of the time are 5 and 15 times more likely to be stable for at least 1~Myr and 10~Myr respectively than systems where this angle librates less than 30\% of the time. While certain resonances seem to prevent systems from short-term instabilities, there is no indication that spending more time in resonance or achieving resonance lock improves stability at the 1~Gyr level. This perhaps is due to the fact we did not find systems with the four planets locked in resonance, which could have improved longer term stability. 

\section{Conclusion}
This paper has aimed to explore the dynamically stable orbits of the HR 8799 system. In the first part of the paper, we continued to demonstrate the precise astrometry that can be achieved by GPI and explored various assumptions on the orbits of the planets.
\begin{itemize}
\item Using GPI IFS data from 2014-2016, we measured the astrometry of the HR 8799 with one milliarcsecond precision using the open-source \texttt{pyKLIP} package.

\item We utilized MCMC methods to fully explore the 20+ dimensional space of orbital configurations using Bayesian parameter estimation and different assumptions on the coplanarity and resonant nature of the system.

\item We found that assuming the system is coplanar or the system is near the 1:2:4:8 period ratio resonance does not significantly worsen the fit to the data, and in fact might make it better. We find including both assumptions provide adequate fits to the data, agreeing well with the conclusions of \citet{Konopacky2016}. 
\end{itemize}

In the second half of this paper, we have presented the first attempt to rigorously fold in dynamical constraints into orbit fits of a directly imaged system, and demonstrated the power of including a dynamical prior.
\begin{itemize}
\item We performed rejection sampling on our posteriors of orbit fits to apply our prior of dynamical stability. Using the \texttt{REBOUND} $N$-body integrator, we ran orbits backwards in time for 40~Myr, the age of the system, varying the masses of the planets and looked for the stable orbital configurations. 

\item We find that coplanar orbits near the 1:2:4:8 resonance produces orders of magnitude more stable orbits than any other scenario. We find a few orbits near 1:2:4:8 resonance with some mutual inclinations that are stable, but the inefficiency of finding them makes studying that family of orbits impractical with current astrometric data. 

\item As demonstrated by Figure \ref{fig:sky-proj}, the stable coplanar orbits lie within a small fraction of the allowed orbital space. In this subspace, we find the outer two planets have near zero eccentricity, while the inner two planets have $e \sim 0.1$. 

\item Our orbits are consistent with being coplanar with the \textit{Herschel}-derived debris disk plane, but
misaligned with the plane derived from SMA and ALMA by $16^{+22}_{-11}$ degrees. Our fitted orbit for planet b is consistent with that planet
sculpting the inner edge of the debris disk in the millimeter,
assuming the orbits are close enough to coplanar.

\item If $M_{cde} \gtrsim6$~$M_{\rm{Jup}}$, planet e needs to be locked in resonance with planet d in order for the system to be stable. Likewise, if $M_{cde} \gtrsim 7$~$M_{\rm{Jup}}$, planet d is likely in resonance lock with planet c. Although we find stable configurations where the inner three planets are in a 1:2:4 Laplace resonance, such a 3-body resonance is not required, as we found many stable configurations where only pairs of planets are in resonance. Planet b does not need to be in resonance for this system to be stable so far. 

\item Using a uniform prior on $M_{cde}$ and a slightly low-mass-favored prior on $M_{b}$, we find 99.9\% of our stable orbits have $M_{b} < 6.3~M_{\rm{Jup}}$ and $M_{cde} < 7.6~M_{\rm{Jup}}$. Folding in the mass constraints from the planet luminosities, hot-start evolutionary models, and a system age of $42 \pm 5$~Myr, we find $M_{b} = 5.8 \pm 0.5~M_{\rm{Jup}}$ and $M_{cde} = 7.2_{-0.7}^{+0.6}~M_{\rm{Jup}}$. Either way, our mass constraints are consistent with hot-start evolutionary tracks. 
    
\item We do not find systems with the inner planets above $7~M_{\rm{Jup}}$ that are stable for the next 1~Gyr. 

\end{itemize}

In the future, as more of the orbital arcs are traced out with precise astrometry, it will become clearer where in the 20+ dimensional space the planets' true orbital configuration lie. In the meantime using more computational power, we can attempt search for stable orbits that are not forced to be exactly coplanar, are stable at higher masses, or are stable at other resonances. Additionally, the new parallax value from Gaia Data Release 2 \citep{Gaia2018} should help tighten the constraints on the semi-major axes of the planets and total mass of the system for the stable coplanar solutions. Even a decade since its discovery, the HR 8799 planetary system is one of the most unique and interesting systems that we know, and combining both detailed dynamical studies with atmospheric characterization will help us understand how these planets formed and how they will interact. 

Our technique of performing rejection sampling to apply dynamical constraints after MCMC sampling of the orbital parameters can also be applied to other directly-imaged multi-planet systems to better constrain the orbits with just a short orbital arc. Here we have shown that even with $<15\%$ of the full orbit covered, we can constrain orbital parameters to a few percent. This can remove the orbital uncertainty that comes with exoplanets discovered through imaging alone, where we typically need to wait for long-period planets to complete a orbital revolution before fully-constraining its orbit. This is also potentially valuable for future space-based imaging missions that search for exo-Earths in multi-planet systems, since this method can estimate the mass and orbit with a small orbital arc. This can allow for the mission to better prioritize which exo-Earth candidates are observed with expensive spectroscopic observations by determining which are most likely Earth-mass and orbiting at a favorable distance from the star. 

\acknowledgements
Simulations in this paper made use of the REBOUND code which can be downloaded freely at http://github.com/hannorein/rebound. We thank Dan Tamayo for offering REBOUND tutorials and help with setting up the simulations. We also thank Eve Lee for helpful discussions on dynamics. 
J.J.W., J.R.G., P.K., and R.J.D.R. were supported in part by NSF AST-1518332, NASA NNX15AC89G and NNX15AD95G. This work benefited from NASA's Nexus for Exoplanet System Science (NExSS) research coordination network sponsored by NASA's Science Mission Directorate. R.I.D. gratefully acknowledges support from NASA XRP 80NSSC18K0355. The Center for Exoplanets and Habitable Worlds is supported by the Pennsylvania State University, the Eberly College of Science, and the Pennsylvania Space Grant Consortium. Portions of this work were performed under the auspices of the U.S. Department of Energy by Lawrence Livermore National Laboratory under Contract DE-AC52-07NA27344.

The GPI project has been supported by Gemini Observatory, which is operated by AURA, Inc., under a cooperative agreement with the NSF on behalf of the Gemini partnership: the NSF (USA), the National Research Council (Canada), CONICYT (Chile), the Australian Research Council (Australia), MCTI (Brazil) and MINCYT (Argentina).
The Gemini Observatory is operated by the Association of Universities for Research in Astronomy, Inc., under a cooperative agreement with the NSF on behalf of the Gemini partnership: the National Science Foundation (United States), the National Research Council (Canada), CONICYT (Chile), the Australian Research Council (Australia), Minist\'erio da Ci\'encia, Tecnologia e Inova\c{c}\~{a}o (Brazil), and Ministerio de Ciencia, Tecnolog\'ia e Innovaci\'on Productiva (Argentina). 
This research has made use of the SIMBAD database, operated at CDS, Strasbourg, France.

\software{Gemini Planet Imager Data Reduction Pipeline \citep{Perrin2014,Perrin2016}, pyKLIP \citep{Wang2015}, emcee \citep{ForemanMackey2013}, REBOUND \citep{Rein2012}, Astropy \citep{Astropy2018}, matplotlib \citep{Hunter2007} }

\facility{Gemini:South (GPI)}

\end{document}